\documentclass[a4paper,11pt]{article}
\usepackage{jheppub}
\usepackage{mathrsfs}
\usepackage{verbatim}
\usepackage{booktabs}
\usepackage{placeins}

\newcommand{\R}{\mathbb{R}}

\newcommand{\C}{\mathbb{C}}

\newcommand{\nn}{\nonumber}
\newcommand{\mc}[1]{\mathcal{#1}}

\newcommand{\pd}[2]{\frac{\partial #1}{\partial #2}}
\newcommand{\tree}{\mathrm{tree}}
\newcommand{\avg}[1]{\langle#1\rangle}
\newcommand{\Avg}[1]{[#1]}
\newcommand{\aAvg}[3]{\langle#1|#2|#3]}

\newcommand{\expval}[3]{\langle #1|\hspace*{.2mm}#2\hspace*{.3mm}|#3 \rangle}
\newcommand{\Res}{\mathop{\rm Res}}
\newcommand{\IP}{\mc I^{\mathrm{P}}}
\newcommand{\INP}{\mc I^{\mathrm{NP}}}
\newcommand{\sign}{\operatorname{sgn}}
\newcommand{\Tr}{\operatorname{Tr}}

\newcommand{\bc}{\begin{center}}
\newcommand{\ec}{\end{center}}

\newcommand{\LRa}{\Longrightarrow}

\newcommand{\LLra}{\Longleftrightarrow}

\title{Global Residues and Two-Loop Hepta-Cuts}

\author{Mads S{\o}gaard}
\affiliation{
Niels Bohr International Academy and Discovery Center, Niels Bohr Institute, \\
University of Copenhagen, Blegdamsvej 17, DK-2100 Copenhagen, Denmark
}

\emailAdd{madss@nbi.dk}

\abstract{We examine maximal unitarity in the nonplanar case and derive 
remarkably compact analytic expressions for coefficients of master integrals with 
two-loop crossed box topology in massless four-point amplitudes in any gauge
theory, thereby providing additional steps towards automated computation of the 
full amplitude. The coefficients are obtained by assembling residues extracted 
through integration on linear combinations of higher-dimensional tori encircling 
global poles of the loop integrand. We recover all salient features of two-loop 
maximal unitarity, such as the existence of unique projectors for each master 
integral. Several explicit calculations are provided. We also establish exact 
equivalence of our results and master integral coefficients recently obtained 
via integrand-level reduction in any renormalizable gauge theory.}

\begin{document} 
\maketitle
\flushbottom

\clearpage
\section{Introduction}
The initiation of the Large Hadron Collider (LHC) programme at CERN has spawned 
a new exciting era in experimental high energy physics and generated an acute
demand for precision cross section predictions for scattering of elementary 
particles. Being a hadron collider, LHC experiments are contaminated 
with a large Quantum Chromodynamics (QCD) background. Discovery of signals of
possibly new physics therefore requires a quantitative understanding of all
relevant Standard Model processes which necessarily must be subtracted from 
the observed data. One-loop scattering amplitudes provide Next-to-Leading Order 
(NLO) estimates, while Next-to-Next-to-Leading order (NNLO) corrections from
two loops are needed for a reliable analysis of theoretical uncertainty.
Although NNLO calculations form the upcoming frontier, two-loop amplitudes are
also relevant already at NLO for processes such as production of diphotons and
pairs of electroweak gauge bosons by gluon fusion for which one-loop is the
leading order.

Scattering amplitudes have traditionally been computed perturbatively by 
translating Feynman diagrams into precise mathematical expressions using 
Feynman rules. This approach gives an invaluable view and interpretation of 
interaction of subatomic particles, but it inevitably suffers from explosive 
growth of complexity with multiplicity and order in perturbation theory. Indeed, 
even in simple problems such as two-by-two gluon scattering an inpracticable 
computational bottleneck is quickly reached. The origin of this problem is that 
intermediate states are virtual particles and a vast amount of redundancy is 
needed for compensation. Catalyzed by Wittens formulation of perturbative gauge
theory as a string theory in twistor space \cite{Witten:2003nn}, new efficient 
on-shell methods for computing tree-level amplitudes using only physical 
information rather than off-shell Feynman diagrams have emerged and striking 
simplicity has been revealed. Most important are the Britto-Cachazo-Feng-Witten 
(BCFW) recursion relations \cite{Britto:2004ap,Britto:2005fq} which remarkably 
construct all gauge theory and also gravity trees by means of just the Cauchy 
residue theorem and complex kinematics in three-point amplitudes whose form is 
actually completely fixed by very general arguments such as scaling properties 
under little group transformations.

Powerful techniques for computation of one-loop amplitudes exploiting unitarity 
of the S-matrix were developed from the Cutkosky rules in the early 1990s by 
Bern, Dixon and Kosower \cite{Bern:1994cg,Bern:1994zx} and subsequently studied 
extensively \cite{Bern:1995db,Bern:1997sc,Britto:2004nc,Britto:2004nj,
Bern:2005hh,Bidder:2005ri,Britto:2005ha,Britto:2006sj,Mastrolia:2006ki,
Brandhuber:2005jw,Ossola:2006us,Anastasiou:2006gt,Bern:2007dw,Forde:2007mi,
Badger:2008cm,Giele:2008ve,Britto:2006fc,Britto:2007tt,Bern:2010qa,
Anastasiou:2006jv}. Unitarity implies that the discontinuity of the transition 
matrix can expressed in terms of simpler quantities, e.g. trees are recycled 
for loops. The unitarity method in its original form allows reconstruction of 
amplitudes from two-particle unitarity cuts that put internal propagators on 
their mass-shell and constrain parameters in an appropriate ansatz. It has 
proven extremely useful in a widespread of both theoretical and phenomenological 
applications in the last two decades, in particular when a proper integral 
basis of the amplitude is not available. The immediate disadvantage is the need 
for performing algebra at intermediate stages because many contributions share 
the same cuts. Generalized unitarity in turn probes the analytic structure of a 
loop integrand much more deeply by imposing several simultaneous on-shell 
conditions, thereby rendering selection of single integrals in a basis 
possible. For instance quadruple cuts isolate a unique box integral
\cite{Britto:2004nc}, whereas other clever projections single out triangles and 
bubbles separately \cite{Forde:2007mi}, leading to beautifully compact 
expressions whose simplicity is by no means expected from a Feynman diagram 
perspective. This method is now fully systematized at one loop with a variety of 
software libraries of numerical implementations that are vital to 
phenomenology at the LHC \cite{Ellis:2007br,Berger:2008sj,Ossola:2007ax,
Mastrolia:2008jb,Giele:2008bc,Berger:2009zg,Badger:2010nx,Berger:2010zx,
Hirschi:2011pa}.

Using current state-of-the-art unitarity techniques one has been able to compute 
four-particle processes in massless QCD \cite{Bern:1997nh,Bern:2000dn,
Glover:2001af,Bern:2002tk,Anastasiou:2000kg,Anastasiou:2000ue,Anastasiou:2001sv}.
It is of obvious interest to extend procedures for direct extraction of integral
coefficients by generalized unitarity beyond one loop. Indeed, it would be of 
enormous theoretical and practical value to have closed form expressions for 
integral coefficients for any two-loop topology such as for instance nonplanar 
crossed double-triangle, planar penta-bubble and planar sunset. Octa-cuts and 
hepta-cuts of two-loop amplitudes in maximally supersymmetric $\mc N = 4$ super 
Yang-Mills theory were first studied in \cite{Buchbinder:2005wp,Cachazo:2008vp}. 
The major obstacle is however that a complete unitarity compatible integral 
basis for two-loop amplitudes is not yet known. On the contrary to one-loop 
integrals whose numerators are trivial, integral basis elements at two loop 
contain complicated tensors. This problem was recently addressed and steps 
towards a solution in that direction were taken in 
\cite{Gluza:2010ws,Schabinger:2011dz}. Although rather technically complicated, 
a very interesting method for obtaining planar double box contributions to 
two-loop amplitudes in any gauge theory using maximal unitarity (i.e. all 
propagators are placed on-shell) cuts has been reported in 
\cite{Kosower:2011ty} and subsequently enhanced and applied in 
\cite{CaronHuot:2012ab,Johansson:2012zv,Johansson:2012sf,Larsen:2012sx}. The
motivation of our paper is to use this framework to analyze nonplanar amplitude
contributions. The continued hope raised by advances along these lines is that 
scattering amplitudes will generate more fundamental insight in hidden structures 
underlying quantum field theories.

The above considerations and the remaining part of this paper resemble a 
perhaps slightly exaggerated, nevertheless quite true, quote by Julian Schwinger: 
{\it one of the most remarkable discoveries in elementary particle physics has 
been that of the existence of the complex plane.} 

\subsection{Conventions and Notation}
\begin{figure}[!h]
\bc
\includegraphics[scale=0.6]{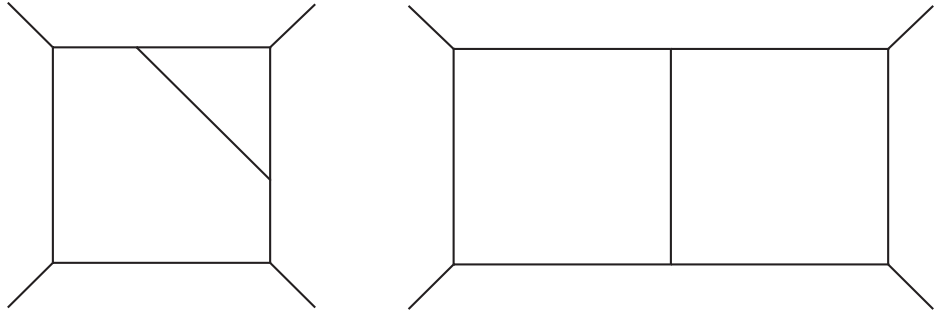}
\caption{The pentabox-triangle and planar double box topologies appearing in the
color-decomposition of the two-loop four-point amplitude.}
\label{TWOLOOPTOPS}
\ec
\end{figure}

In this paper we consider color-ordered scattering amplitudes at two-loops in 
gauge theory with $SU(N_c)$ symmetry group in which case decoupling of color and
kinematical structures is also important like at tree-level and one-loop. The 
color-dressed two-loop amplitude with four external particles transforming in 
the adjoint representation of the gauge group admits color decomposition in 
terms of single and double traces,
\begin{align}
\mathbb{A}_4^{\text{2-loop}} =
\sum_{\sigma\in S_4/Z_4^3} 
N_c\Tr(T^{a_{\sigma(1)}}T^{a_{\sigma(2)}})
\Tr(T^{a_{\sigma(3)}}T^{a_{\sigma(4)}})
A^{(2)}_{4;1,3}(\sigma(1),\sigma(2);\sigma(3),\sigma(4))\;\;\; \nn \\[1mm]
+\sum_{\sigma\in S_4/Z_4} 
\Tr(T^{a_{\sigma(1)}}T^{a_{\sigma(2)}}T^{a_{\sigma(3)}}T^{a_{\sigma(4)}})
\Big[ 
N_c^2 A^{(2),LC}_{4;1,1}(\sigma(1),\sigma(2),\sigma(3),\sigma(4))
\nn\;\;\; \\[1mm]
+A^{(2),SC}_{4;1,1}(\sigma(1),\sigma(2),\sigma(3),\sigma(4)) 
\Big]\;,
\end{align}
where $T^a$ for $a = 1,\dots,N_c^2-1$ are generators of $SU(N_C)$ in the
fundamental representation. The color-stripped amplitudes on the right hand side 
all have expansions as linear combinations of integrals such as the planar
double box, nonplanar crossed box and triangle-pentabox (see 
fig.~\ref{TWOLOOPTOPS}) with legs permuted appropriately. The complete map is 
excluded here for brevity, but available in \cite{Badger:2012dp}. In this form, 
color-ordered generalized unitarity cuts can be applied. 

Partial amplitudes are naturally built from antisymmetric Lorentz invariant 
holomorphic and antiholomorphic inner products of commuting spinors 
$\lambda_i^\alpha$ and $\tilde\lambda_i^{\dot\alpha}$ whose components are 
homogeneous coordinates on complex projective space $\mathbb{CP}^1$. Physically, 
the spinors are solutions of definite chirality to the massless Dirac equation. 
We define angle and square brackets by
\begin{align}
\avg{ij} = -\avg{ji} \equiv 
\epsilon_{\alpha\beta}\lambda_i^\alpha\lambda_j^\beta\;, \quad
\Avg{ij} = -\Avg{ji} \equiv 
\epsilon_{\dot\alpha\dot\beta}
\tilde\lambda_i^{\dot\alpha}\tilde\lambda_j^{\dot\beta}
\end{align}
and identify the corresponding four-dimensional null-momentum 
$k_i^{\alpha\dot\alpha} = \lambda_i^\alpha\lambda_i^{\dot\alpha}$. Frequently
used momentum invariants can then be written
\begin{align}
s_{ij} = \avg{ij}\Avg{ji} = 2k_i\cdot k_j
\end{align}
with Mandelstam variables $s\equiv s_{12}$, $u\equiv s_{13}$ and 
$t\equiv s_{14}$ such that $s+t+u = 0$. Momenta are by convention outgoing and 
summed using the notation $K_{i_1\cdots i_n} = k_{i_1}+\cdots+k_{i_n}$. Our 
expressions also involve parity-odd contractions between Levi-Civita symbols 
and momenta in the form
\begin{align}
\varepsilon(1,2,3,4) = {} &
\sum_{\sigma\in Z_4}(\sign \sigma)
k_{1,\sigma(1)} k_{2,\sigma(2)} k_{3,\sigma(3)} k_{4,\sigma(4)} 
\nn \\ = {} & \,
\frac{i}{4}(
\avg{12}\Avg{23}\avg{34}\Avg{41}-
\Avg{12}\avg{23}\Avg{34}\avg{41})\;.
\end{align}

\section{Generalized Unitarity and Integral Bases}
The existence of a finite basis of linearly independent scalar integrals 
for one-loop gauge theory amplitudes has in recent years established a solid 
foundation for the success of the modern formulation of the unitarity method. 
Using Passarino-Veltmann reduction an $n$-point amplitude can be written as
\begin{align}
\label{ONELOOPMASTER}
A_n^{\text{1-loop}} = \sum_{\text{boxes}}c_\square I_\square+
\sum_{\text{triangles}}c_\triangle I_\triangle+
\sum_{\text{bubbles}}c_\circ I_\circ+
\sum_{\text{tadpoles}}c_{-\hspace*{-0.5mm}\circ}
I_{-\hspace*{-0.5mm}\circ}+\text{rational terms}\;,
\end{align}
where scalar bubble, triangle and box integrals are known in dimensional
regularization explicitly and tadpoles are present only in case of massive
internal propagators. In a nutshell, computation of one-loop amplitudes is thus 
reduced to finding the rational coefficients in the integral basis. At one-loop, 
direct extraction procedures exist for all topologies \cite{Forde:2007mi} and 
even for the rational terms \cite{Badger:2008cm}.

In this section we describe an approach to maximal unitarity introduced in
\cite{Buchbinder:2005wp,Cachazo:2008vp} and recently systematized for general 
planar double boxes in \cite{Kosower:2011ty,CaronHuot:2012ab} using unitarity
compatible integral bases and complex analysis in higher dimensions.

\subsection{Multivariate Residue Theorem}
The extension of the one-dimensional version of the Cauchy residue theorem to 
several complex variables has proven advantageous in order to understand
computations of generalized unitarity cuts of multiloop amplitudes. We therefore
now introduce the concept of global poles and the global residue theorem, and 
refer the reader to \cite{ArkaniHamed:2009dn} for further information.

Let the meromorphic function $\varphi:\C^2\to\C$ be given by
\begin{align}
\varphi(z_1,z_2) = \frac{h(z_1,z_2)}{(az_1+bz_2+c)(ez_1+fz_2+g)}\;,
\end{align}
and assume regularity of $h(z_1,z_2)$ where the denominators vanish simultaneously, 
that is ${(az_1+bz_2+c) = 0}$ and ${(ez_1+fz_2+g) = 0}$. Such a point 
${(z_1^\star,z_2^\star)\in\C^2}$ is called a global pole for $\varphi$. Then we can 
consider the multidimensional contour integral of $\varphi$ on an infinitesimal 
two-torus $T_\epsilon^2 \simeq S^1\times S^1$ encircling that global pole. 
Moreover, we can shift the global pole to origo by applying the change of variables 
${w_1 = az_1+bz_2+c}$ and ${w_2 = ez_1+fz_2+g}$,
\begin{align}
\oint_{T_\epsilon^2(z_1^\star,z_2^\star)}
\frac{h(z_1,z_2)dz_1dz_2}{(az_1+bz_2+c)(ez_1+fz_2+g)} = 
\oint_{T_\epsilon^2(0,0)}\frac{dw_1dw_2}{w_1w_2}\frac{h(z_1(w),z_2(w))}{
\det\left(\frac{\partial(w_1,w_2)}{\partial(z_1,z_2)}\right)}\;,
\end{align}
whence in analogy with the one-dimensional case it is very natural to define the 
global residue of $\varphi$ at $(z_1^\star,z_2^\star)$ by
\begin{align}
\Res_{(z_1,z_2)=(z_1^\star,z_2^\star)} f(z_1,z_2) =
\frac{h(z_1^\star,z_2^\star)}{\det\left(\frac{
\partial(w_1,w_2)}{\partial(z_1,z_2)}\right)\!\Big|_{(z_1^\star,z_2^\star)}}\;.
\end{align}
The generalization to meromorphic functions $\varphi:\C^n\to\C$ of $n$ complex 
variables and with $m\geq n$ factors in the denominator,
\begin{align}
\varphi(z_1,\dots,z_n) = \frac{h(z_1,\dots,z_n)}
{\prod_{i=1}^m p_i(z_1,\dots,z_n)}\;,
\end{align}
is straightforward. Indeed, we solve 
$p_{i_1}(z_1^\star,\dots,z_n^\star) = \cdots = 
p_{i_n}(z_1^\star,\dots,z_n^\star) = 0$ to determine the global pole 
$z^\star = (z_1^\star,\dots,z_n^\star)\in\C^n$. By assumption $h$ is regular there 
and the global residue of $\varphi$ thus reads
\begin{align}
\oint_{T_\epsilon^n(z^\star)}d^nz
\frac{h(z_1,\dots,z_n)}
{\prod_{i=1}^m p_i(z_1,\dots,z_n)} = 
\frac{h(z_1^\star,\dots,z_n^\star)}
{\prod_{i\neq(i_1,\dots,i_n)}p_i(z_1^\star,\dots,z_n^\star)
\det\left(\frac{\partial(p_{i_1},\dots,p_{i_n}}
{\partial(z_1,\dots,z_n)}\right)\!\Big|_{(z_1^\star,\dots,z_n^\star)}}\;.
\end{align}

In this way, actually ${m\choose n}$ global residues arise. From now on we will
only encounter situations where $n = m$ so that the integral localizes to a
single residue. 

Strictly speaking, in order for the global residue to become independent of the
orientation of the parametrization of the torus, the integration variables should 
really be wedged together. However, this point is irrelevant for our purposes as
long as the orientation is kept consistent throughout the entire calculation.

\subsection{Method of Maximal Cuts}
\begin{figure}[!h]
\bc
\includegraphics[scale=0.75]{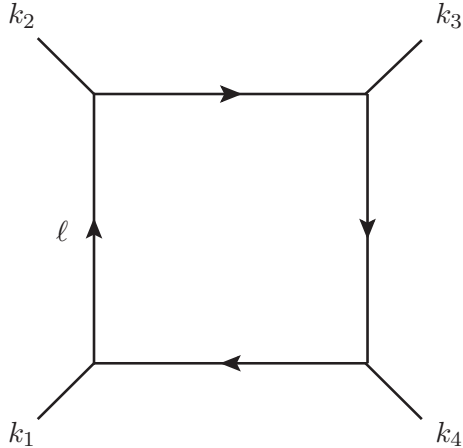}
\put(-160,-5){$k_1$}
\put(-160,153){$k_2$}
\put(0,153){$k_3$}
\put(0,-5){$k_4$}
\put(-142,71){$\ell$}
\caption{The four-point massless one-loop box diagram.}
\label{ONELOOPBOX}
\ec
\end{figure}

Let us return to the application to generalized unitarity and focus our attention 
on extraction of the coefficient in front of the four-point one-loop scalar box 
integral (fig.~\ref{ONELOOPBOX})
\begin{align}
\mc I_\square(s,t) \equiv
\int_{\R^D}\!\frac{d^D\ell}{(2\pi)^D}
\frac{1}{\ell^2(\ell-k_2)^2(\ell-K_{23})^2(\ell+k_1)^2}\;,
\end{align}
with external momenta $k_1,\dots,k_4$. For each such quartet of momenta the 
solution set $\mc S$ for the quadruple cut equations formed from the zero locus 
of the four inverse propagators is a pair of complex
conjugates\footnote{Technically speaking, identification by complex conjugation
presumes reality of momenta.} $\mc S_1$ and $\mc S_2$,
\begin{align}
\mc S = 
\big\{\ell\in\C^4\;|\; &
\ell^2 = 0\,,\;
(\ell-k_2)^2 = 0\,,\;
(\ell-K_{23})^2 = 0\,,\;
(\ell+k_1)^2 = 0\big\} = 
\mc S_1\cup\mc S_2\;.
\end{align}
The kinematical structure of the solutions is easy to understand since they 
correspond to the two possible configurations of nonconsecutive holomorphically 
and antiholomorphically collinear three-vertices in a box.

We now adopt the ideas of \cite{Britto:2004nc}, later clarified in 
\cite{Kosower:2011ty}, and define the quadruple cut of a general box integral by 
shifting integration region from $\R^4$ to a surface embedded in $\C^4$ formed by 
a linear combination of the two four-tori encircling the leading singularities 
$\mc S_1$ and $\mc S_2$,
\begin{align}
\int_{\R^D}\frac{d^D\ell}{(2\pi)^D}\frac{\mc P(\ell)}
{\prod_{k=1}^4p_k^2(\ell)}\;\xrightarrow{\text{cut}}\;
\sum_{i=1,2}\Lambda_i\oint_{T_i}\frac{d^4\ell}{(2\pi)^4}
\frac{\mc P(\ell)}{\prod_{k=1}^4p_k^2(\ell)}\;.
\end{align}
Notice that we always strip all expected occurrences of factors of $2\pi i$. The 
contour weights or winding numbers $\Lambda_1$ and $\Lambda_2$ are a priori 
unknown, but consistency constraints from integral reduction fix their relative 
normalization to unity. Applying this recipe to both sides of the master integral 
equation \eqref{ONELOOPMASTER} we obtain the augmented quadruple cut
\begin{align}
c_\square\sum_{i = 1,2}
\oint_{T_i}\!\frac{d^4\alpha}{(2\pi)^4}
\left(\det_{\mu,j}\pd{\ell^\mu}{\alpha_j}\right)
\prod_{k=1}^4\frac{1}{p_k^2(\alpha)} \hspace*{7cm} \nn \\ 
= \sum_{i = 1,2}
\sum_{\substack{\text{helicities}\\\text{particles}}}
\oint_{T_i}\!\frac{d^4\alpha}{(2\pi)^4}
\left(\det_{\mu,j}\pd{\ell^\mu}{\alpha_j}\right)
\prod_{k=1}^4
\frac{1}{p_k^2(\alpha)}\tilde A_{(k)}^\tree(\alpha)\;,
\end{align}
where we absorbed the contour weights into the integrals and also put a tilde on 
the tree amplitudes to indicate that they are really off-shell until the contour 
integral is localized onto the cut solutions. Linearity of the loop momentum in 
$\alpha_1,\dots,\alpha_4$ implies that the Jacobian is constant and therefore it 
can be ignored. We can also cancel common factors on both sides and discard the 
Jacobian arising from actually evaluating the contour integrals in parameter space 
and obtain the well-known Britto-Cachazo-Feng formula \cite{Britto:2004nc}
\begin{align}
c_\square = \frac{1}{2}
\sum_{i=1,2}
\sum_{\substack{\text{helicities}\\ \text{particles}}}
\prod_{k=1}^4 A_{(k)}^\tree\big|_{\mc S_i}\;.
\end{align}
Strikingly simple, it singles out uniquely any one-loop gauge theory scalar box 
integral coefficient in terms of just a product of four tree amplitudes evaluated 
at complex momenta arising by promoting all internal lines to on-shell values.

This approach generalizes to two loops and presumably beyond using the following 
principle \cite{Kosower:2011ty}. We define the maximal cut by continuation of real 
slice $L$-loop integrals into $(\C^4)^{\otimes L}$ by choosing contours that 
encircle the true global poles of the integrand in such a way that any integral 
identity in $(\R^D)^{\otimes L}$ is preserved. If necessary, impose auxiliary cut 
constraints by localizing remaining integrations onto composite leading 
singularities or poles in tensor integrands to obtain linear algebraic equations 
that uniquely determine the master integral coefficients from tree-level data.

\begin{figure}[!h]
\bc
\includegraphics[scale=0.8]{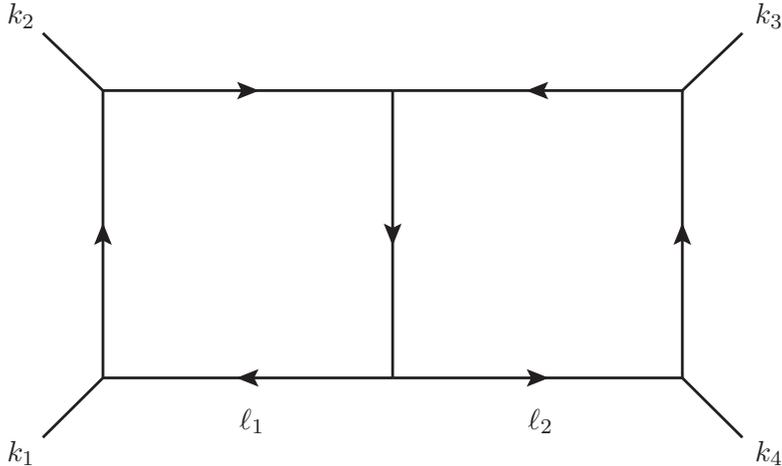}
\put(-280,-5){$k_1$}
\put(-280,160){$k_2$}
\put(0,160){$k_3$}
\put(0,-5){$k_4$}
\put(-193,7){$\ell_1$}
\put(-85,7){$\ell_2$}
\caption{The massless four-point planar double box diagram. External momenta are 
by convention taken as outgoing.}
\label{DBOXDIAGRAM}
\ec
\end{figure}

Consider in brevity the application of this prescription to the primitive 
amplitude for the four-point planar double box with massless kinematics. The 
Feynman integral for the diagram shown in fig.~\ref{DBOXDIAGRAM} reads
\begin{align}
\IP[1] \equiv
\int_{\R^D}\!\frac{d^D\ell_1}{(2\pi)^D}
\int_{\R^D}\!\frac{d^D\ell_2}{(2\pi)^D}
\frac{1}{\ell_1^2(\ell_1-k_1)^2(\ell_1-K_{12})^2\ell_2^2
(\ell_2-k_4)^2(\ell_2-K_{34})^2(\ell_1+\ell_2)^2}\;.
\end{align}
In general, the integral may have an arbitrary numerator and in that case we
write $\IP[\mc P(\ell_1,\ell_2)]$. Integrals of this type were calculated 
analytically in \cite{Smirnov:1999gc,Smirnov:1999wz}.

It is now easy to write down and solve the seven on-shell constraints in 
parameter space using the same parametrization of the loop momenta as for the 
nonplanar double box below \eqref{LOOPPARAM}. Each solution has a free complex 
parameter $z$ that parametrizes a Riemann surface of genus 0. Direct evaluation 
reveals that the localization of the double box scalar integral onto this 
remaining Riemann sphere yields the same Jacobian for all six solutions, with the 
very simple result
\begin{align}
\IP[1]_{\mc S_i} = -\frac{1}{16s_{12}^3}\oint\frac{dz}{z(z+\chi)}\;.
\end{align}
We impose an eighth cut condition and freeze the remaining integral completely by
choosing linear combinations of contours encircling the Jacobian poles 
$z\in\{0,-\chi\}$ and additional tensor poles at $z = -\chi-1$ in integrals with
nontrivial numerators. In total we naively find fourteen candidate global poles.

By virtue of integration-by-parts identities among renormalizable Feynman
integrals, the double box primitive amplitude may be expanded in an integral 
basis whose elements are, for instance, $\IP[1]$ and $\IP[(\ell_1\cdot k_4)]$,
\begin{align}
A^{\text{2-loop}}_{\text{dbox}} = 
c_1\,\IP[1]+c_2\,\IP[(\ell_1\cdot k_4)]+\cdots\;.
\end{align}
Integrals with subleading topologies are hidden in the ellipses. All
seven-propagator integration-by-parts identities are available in
appendix~\ref{IBPDBOX}. The augmented hepta-cut of the master integral equation 
may then be derived from residue identities between on-shell branches and 
identification of eight true global poles along the lines of 
\cite{CaronHuot:2012ab}. In particular, the double box primitive amplitude 
factorizes onto a product of six tree-level amplitudes arranged in six distinct 
configurations such that no external legs are neither holomorphically nor 
antiholomorphically collinear for generic momenta. 

Requiring that all reduction identities continue to hold after imposing the 
hepta-cut constraints leads to unique projectors for the two master integral 
coefficients, up to an irrelevant overall normalization. Following the enumeration 
of on-shell solutions in \cite{Kosower:2011ty}, one possible minimal 
representation is the residue expansion
\begin{align}
\label{DBOXMASTER1}
c_1 = {} & +\frac{1}{4}\sum_{i=1,3}\Res_{z=-\chi}\frac{1}{z+\chi}
\sum_{\substack{\text{particles}\\ \text{helicities}}}
\prod_{j=1}^6 A_{(j)}^\tree(z)\big|_{\mc S_i} \nn \\ &
+\frac{1}{4}\sum_{i=5,6}\Res_{z=-\chi}\frac{1}{z+\chi}
\sum_{\substack{\text{particles}\\ \text{helicities}}}
\prod_{j=1}^6 A_{(j)}^\tree(z)\big|_{\mc S_i} \nn \\ &
-\frac{\chi}{4(1+\chi)}\sum_{i=5,6}\Res_{z=-\chi-1}
\sum_{\substack{\text{particles}\\ \text{helicities}}}
\prod_{j=1}^6 A_{(j)}^\tree(z)\big|_{\mc S_i}\;, \\[2mm]
\label{DBOXMASTER2}
c_2 = {} &
-\frac{1}{2s_{12}\chi}\sum_{i=1,3}
\Res_{z=-\chi}\frac{1}{z+\chi}
\sum_{\substack{\text{particles}\\ \text{helicities}}}
\prod_{j=1}^6 A_{(j)}^\tree(z)\big|_{\mc S_i} \nn \\ &
+\frac{1}{s_{12}\chi}\sum_{i=5,6}
\Res_{z=0}\frac{1}{z}
\sum_{\substack{\text{particles}\\ \text{helicities}}}
\prod_{j=1}^6 A_{(j)}^\tree(z)\big|_{\mc S_i} \nn \\ &
-\frac{1}{2s_{12}\chi}\sum_{i=5,6}
\Res_{z=-\chi}\frac{1}{z+\chi}
\sum_{\substack{\text{particles}\\ \text{helicities}}}
\prod_{j=1}^6 A_{(j)}^\tree(z)\big|_{\mc S_i} \nn \\ &
+\frac{3}{2s_{12}(1+\chi)}\sum_{i=5,6}
\Res_{z=-\chi-1}
\sum_{\substack{\text{particles}\\ \text{helicities}}}
\prod_{j=1}^6 A_{(j)}^\tree(z)\big|_{\mc S_i}\;,
\end{align}
in which on-shell branches $S_2$ and $S_4$ are eliminated.

\section{Nonplanar Crossed Box}
Conventional wisdom and numerous experiences suggest that nonplanar diagrams in
general are more complicated to compute than planar ones. In this section we 
provide additional evidence in favor of the approach to maximal unitarity
described above by revealing surprising simplicity in the nonplanar crossed box. 
In particular, we establish the augmented hepta-cut and derive beautiful formulae 
for the master integral coefficients from unique projectors, highlighting 
differences and similarities to the planar double box in the process.

\begin{figure}[!h]
\bc
\includegraphics[scale=0.8]{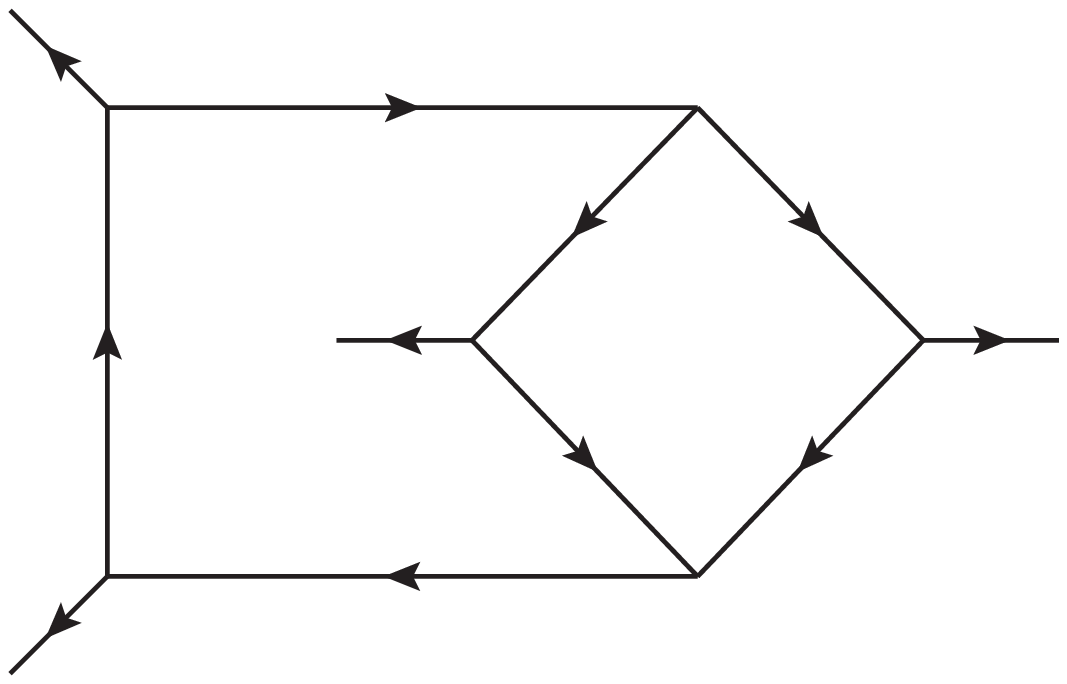}
\put(0,79){$k_1$}
\put(-188,79){$k_2$}
\put(-265,0){$k_3$}
\put(-265,157){$k_4$}
\put(-245,78){$\ell_2$}
\put(-52,47){$\ell_1$}
\caption{Momentum flow for the twoloop crossed box.}
\label{XBOXDIAGRAM}
\ec
\end{figure}

The dimensionally regularized Feynman integral for the four-point nonplanar double 
box with massless kinematics and an arbitrary numerator function 
$\mc P(\ell_1,\ell_2)$ inserted is
\begin{align}
\INP[\mc P(\ell_1,\ell_2)] \equiv
\int_{\R^D}\!\frac{d^D\ell_1}{(2\pi)^D}
\int_{\R^D}\!\frac{d^D\ell_2}{(2\pi)^D}
\frac{\mc P(\ell_1,\ell_2)}{
\ell_1^2(\ell_1+k_1)^2\ell_2^2(\ell_2+k_3)^2} \nn \hspace*{3cm}\, \\[1mm]\
\times \frac{1}{(\ell_2-k_4)^2
(\ell_2-\ell_1+k_3)^2(\ell_2-\ell_1+K_{23})^2}\;,
\end{align}
following the conventions outlined in fig.~\ref{XBOXDIAGRAM}. In a slight abuse
of terminology it is called a tensor integral even though it has no free
indices. Explicit expressions for these integrals are available in 
\cite{Tausk:1999vh,Anastasiou:2000mf}.

\subsection{Parametrization of On-Shell Solutions}
In order to study the hepta-cut, we exploit slight calculational foresight and 
choose convenient normalizations in the parametrization of the two independent 
loop momenta,
\begin{align}
\label{LOOPPARAM}
\ell_1^\mu(\alpha_1,\dots,\alpha_4) & = 
\alpha_1k_1^\mu+\alpha_2k_2^\mu+
\frac{s_{12}\alpha_3}{2\avg{14}\Avg{42}}\expval{1^-}{\gamma^\mu}{2^-}+
\frac{s_{12}\alpha_4}{2\avg{24}\Avg{41}}\expval{2^-}{\gamma^\mu}{1^-}\;, \\[1mm]
\ell_2^\mu(\beta_1,\dots,\beta_4) & = 
\beta_1k_3^\mu+\beta_2k_4^\mu+
\frac{s_{12}\beta_3}{2\avg{31}\Avg{14}}\expval{3^-}{\gamma^\mu}{4^-}+
\frac{s_{12}\beta_4}{2\avg{41}\Avg{13}}\expval{4^-}{\gamma^\mu}{3^-}\;.
\end{align}
The virtue of this form is maximal simplification of the hepta-cut equations and
direct exposure of global residues of the integrand. The Jacobians for the
change of variables from momenta to parameters are constant and can therefore be
disregarded in the augmented hepta-cut below, but for completeness we note that
\begin{align}
\label{LOOPPARAMETERJACOBIAN}
J_\alpha = \det_{\mu,i}\pd{\ell_1^\mu}{\alpha_i} =
-\frac{is_{12}^2}{4\chi(\chi+1)}\;, \quad
J_\beta = \det_{\mu,i}\pd{\ell_2^\mu}{\beta_i} = 
-\frac{is_{12}^2}{4\chi(\chi+1)}\;,
\end{align}
where $\chi$ is a ratio of Mandelstam invariants used throughout this
calculation,
\begin{align}
\chi = \frac{s_{14}}{s_{12}}\;.
\end{align}

The on-shell equations are maximally degenerate for the kinematical
configuration in consideration and rather straightforward to analyze. The 
solution set $\mc S$ is the union of eight irreducible branches $\mc S_i$, each 
of which is topologically equivalent to a Riemann sphere,
\begin{align}
\mc S = 
\big\{(\ell_1,\ell_2)\in(\C^4)^{\otimes 2}\;|\; &
\ell_1^2 = 0\,,\;
(\ell_1+k_1)^2 = 0\,,\;
\ell_2^2 = 0\,,\; \nn \\[2mm] & \! 
(\ell_2+k_3)^2\,,\;
(\ell_2+k_3)^2 = 0\,, 
(\ell_2-k_4)^2 = 0\,,\; \nn \\[-1mm] & \!
(\ell_1-\ell_2-k_3)^2 = 0\,,\;
(\ell_1-\ell_2-K_{23})^2 = 0\big\} = 
\bigcup_{i=1}^8 \mc S_i\;.
\end{align}

Let us solve the hepta-cut equations using the parametrization of $\ell_1$ and
$\ell_2$. We examine the subset of inverse propagators involving only a single 
loop momentum on the cut, and obtain
\begin{align}
\ell_1^2 = 
s_{12}\bigg(\alpha_1\alpha_2+
\frac{\alpha_3\alpha_4}{\chi(\chi+1)}\bigg) = 0\;, \nn \\
\ell_2^2 = s_{12}\bigg(\beta_1\beta_2+
\frac{\beta_3\beta_4}{\chi(\chi+1)}\bigg) = 0\;, \nn \\
(\ell_1+k_1)^2 = s_{12}\bigg((\alpha_1+1)\alpha_2+
\frac{\alpha_3\alpha_4}{\chi(\chi+1)}\bigg) = 0\;, \nn \\
(\ell_2+k_3)^2 = s_{12}\bigg((\beta_1+1)\beta_2+
\frac{\beta_3\beta_4}{\chi(\chi+1)}\bigg) = 0\;, \nn \\
(\ell_2-k_4)^2 = s_{12}\bigg(\beta_1(\beta_2-1)+
\frac{\beta_3\beta_4}{\chi(\chi+1)}\bigg) = 0\;.
\end{align}
These constraints translate into $\alpha_2 = \beta_1 = \beta_2 = 0$, 
$\alpha_3\alpha_4 = 0$ and $\beta_3\beta_4 = 0$ for generic kinematics, and
therefore we have to consider four types of solutions. For completeness, we
derive equations for the mixed inverse propagators on the hepta-cut whose form 
is compatible with any kind of solution,
\begin{align}
(\ell_1-\ell_2-k_3)^2\big|_{\text{cut}} = s_{12}\bigg[\,&
\alpha_1(1+\chi-\beta_3-\beta_4)+\alpha_3+\alpha_4 \nn \\ &
-\frac{1}{\chi}(\alpha_3\beta_3+\alpha_4\beta_4) 
-\frac{1}{\chi+1}(\alpha_3\beta_4+\alpha_4\beta_3)
\,\bigg]_{\text{cut}}\;, \\
(\ell_1-\ell_2-K_{2,3})^2\big|_{\text{cut}} = s_{12}\bigg[\,&
\alpha_1(\chi-\beta_3-\beta_4)
-\frac{1}{\chi}(\alpha_3\beta_3+\alpha_4\beta_4) \nn \\ &
-\frac{1}{\chi+1}(\alpha_3\beta_4+\alpha_4\beta_3)
+\alpha_3+\alpha_4-\beta_3-\beta_4+\chi\,\bigg]_{\text{cut}}\;,
\end{align}
where the cut subscript means $\xi\to0$ for
$\xi\in\{(\alpha_3,\beta_3),
(\alpha_3,\beta_4),(\alpha_4,\beta_3),(\alpha_4,\beta_4)\}$. It is trivial to
show that the these hepta-cut equations collapse into two classes; for 
$\alpha_j = \beta_j = 0$ and $i\neq j$,
\begin{align}
(\beta_i-\chi)(\alpha_i+\chi+\alpha_1\chi) = 0\;, \nn \\
\alpha_i(1-\beta_i/\chi)+\alpha_1(1-\beta_i+\chi) = 0\;,
\end{align}
whereas for $\alpha_j = \beta_i = 0$ with $i \neq j$,
\begin{align}
\alpha_1(1-\beta_j+\chi)+\alpha_i
\left(1-\frac{\beta_j}{1+\chi}\right) = 0\;, \nn \\
(1+\alpha_1)(\beta_j-\chi)-\alpha_i\left(1-\frac{\beta_j}{1+\chi}\right) = 0\;.
\end{align}
Each set of equations has again two independent branches, whence upon
parametrization of the remaining freedom by the complex variable $z\in\C$ we 
arrive at the eight solutions listed in table~\ref{NONPLANARSOLUTIONS}. 
The appearance of four pairs of complete conjugates is naturally expected in 
view of the, for generic momenta, valid distributions of internal helicities 
in the six three-vertices on the hepta-cut, see appendix~\ref{CROSSEDBOXTREES}.

\begin{table}
\begin{align}
\begin{array}{@{}ccccccccc@{}}
\toprule
\;\;&\;\; \alpha_1 \;\;&\;\; \alpha_2 \;\;&\;\; \alpha_3 \;\;&\;\; \alpha_4 
\;\;&\;\; \beta_1 \;\;&\;\; \beta_2 \;\;&\;\; \beta_3 \;\;&\;\; \beta_4 \\  
\midrule
\mc S_1 \;\;&\;\; \chi-z \;\;&\;\; 0 \;\;&\;\; \chi(z-\chi-1) \;\;&\;\; 0
\;\;&\;\; 0 \;\;&\;\; 0 \;\;&\;\; z \;\;&\;\; 0 \\
\mc S_2 \;\;&\;\; \chi-z \;\;&\;\; 0 \;\;&\;\; 0 \;\;&\;\; \chi(z-\chi-1) 
\;\;&\;\; 0 \;\;&\;\; 0 \;\;&\;\; 0 \;\;&\;\; z \\
\mc S_3 \;\;&\;\; 0 \;\;&\;\; 0 \;\;&\;\; z \;\;&\;\; 0
\;\;&\;\; 0 \;\;&\;\; 0 \;\;&\;\; \chi \;\;&\;\; 0 \\
\mc S_4 \;\;&\;\; 0 \;\;&\;\; 0 \;\;&\;\; 0 \;\;&\;\; z
\;\;&\;\; 0 \;\;&\;\; 0 \;\;&\;\; 0 \;\;&\;\; \chi \\
\mc S_5 \;\;&\;\; \chi-z \;\;&\;\; 0 \;\;&\;\; 0 \;\;&\;\; (\chi+1)(z-\chi)
\;\;&\;\; 0 \;\;&\;\; 0 \;\;&\;\; z \;\;&\;\; 0 \\
\mc S_6 \;\;&\;\; \chi-z \;\;&\;\; 0 \;\;&\;\; (\chi+1)(z-\chi) \;\;&\;\; 0
\;\;&\;\; 0 \;\;&\;\; 0 \;\;&\;\; 0 \;\;&\;\; z \\
\mc S_7 \;\;&\;\; -1 \;\;&\;\; 0 \;\;&\;\; 0 \;\;&\;\; z
\;\;&\;\; 0 \;\;&\;\; 0 \;\;&\;\; 1+\chi \;\;&\;\; 0 \\
\mc S_8 \;\;&\;\; -1 \;\;&\;\; 0 \;\;&\;\; z \;\;&\;\; 0
\;\;&\;\; 0 \;\;&\;\; 0 \;\;&\;\; 0 \;\;&\;\; 1+\chi \\
\bottomrule
\end{array}
\nn
\end{align}
\caption{The eight solutions to the on-shell equations for the maximal cut of
the four-point massless nonplanar double box. Each irreducible branch has 
topology of a genus-0 sphere.}
\label{NONPLANARSOLUTIONS}
\end{table}

\FloatBarrier
\subsection{Composite Leading Singularities}
Let us now apply the hepta-cut to the nonplanar double box primitive amplitude. 
For each solution to the on-shell equations we have to compute the Jacobian 
associated with the localization of the integral onto a single Riemann sphere. 
We will work out the case appropriate to the first solution in detail. 

Initially we use all constraints involving only either $\ell_1$ or $\ell_2$,
\begin{align}
J_A = {} & \frac{1}{s_{12}^5}
\oint_{C_\epsilon(0)}\!d\alpha_2\oint_{C_\epsilon(0)}\!d\alpha_4\,
\frac{1}{\alpha_1\alpha_2+\frac{\alpha_3\alpha_4}{\chi(\chi+1)}}
\frac{1}{(\alpha_1+1)\alpha_2+\frac{\alpha_3\alpha_4}{\chi(\chi+1)}} \times 
\nn \\[1mm] &
\oint_{C_\epsilon(0)}\!d\beta_1
\oint_{C_\epsilon(0)}\!d\beta_2
\oint_{C_\epsilon(0)}\!d\beta_4\,
\frac{1}{\beta_1\beta_2+\frac{\beta_3\beta_4}{\chi(\chi+1)}}
\frac{1}{(\beta_1+1)\beta_2+\frac{\beta_3\beta_4}{\chi(\chi+1)}}
\frac{1}{\beta_1(\beta_2-1)+\frac{\beta_3\beta_4}{\chi(\chi+1)}}\,,
\end{align}
and then combine with integrals containing both loop momenta on this support,
\begin{align}
J_B = \frac{1}{s_{12}^2}
\oint_{C_\epsilon(\mu)}\!d\alpha_1
\oint_{C_\epsilon(\lambda)}\!d\alpha_3
\frac{1}{(\chi-\beta_3)(1+\alpha_1+\chi^{-1}\alpha_3)}
\frac{1}{\alpha_1(1+\chi-\beta_3)+\alpha_3(1-\chi^{-1}\beta_3)}\;,
\end{align}
where we put $\mu = \chi-\beta_3$ and $\lambda = \chi(\beta_3-\chi-1)$. The
seven contour integrals are evaluated as determinants using the multivariate 
residue theorem and produce the rather simple forms
\begin{gather}
J_A^{-1} = s_{12}^5\det\left(
\begin{array}{cc}
\alpha_1 & \frac{\alpha_3}{\chi(\chi+1)} \\
\alpha_1+1 & \frac{\alpha_3}{\chi(\chi+1)}
\end{array}
\right) 
\det\left(
\begin{array}{ccc}
\beta_2 & \beta_1 & \frac{\beta_3}{\chi(\chi+1)} \\
\beta_2 & \beta_1+1 & \frac{\beta_3}{\chi(\chi+1)} \\
\beta_2-1 & \beta_1 & \frac{\beta_3}{\chi(\chi+1)}
\end{array}
\right) = 
-\frac{s_{12}^5\alpha_3\beta_3}{\chi^2(\chi+1)^2}\;, \\[1mm]
J_B^{-1} = s_{12}^2\det\left(
\begin{array}{cc}
1+\chi-\beta_3 & 1-\frac{\beta_3}{\chi} \\
\chi-\beta_3 & 1-\frac{\beta_3}{\chi}
\end{array}
\right) = s_{12}^2\Big(1-\frac{\beta_3}{\chi}\Big)\;.
\end{gather}
We include previous effects of change of variables 
\eqref{LOOPPARAMETERJACOBIAN} and derive the full Jacobian
\begin{align}
\INP[1]_{\mc S_1} = -\frac{\chi}{16s_{12}^3}\oint
\frac{d\beta_3}{\alpha_3\beta_3(\beta_3-\chi)}\;,
\end{align}
which in the specific parametrization of $\alpha_3$ and $\beta_3$ becomes
\begin{align}
\INP[1]_{\mc S_1} =
-\frac{1}{16s_{12}^3}\oint\frac{dz}{z(z-\chi)(z-\chi-1)}\;.
\end{align}

The remaining seven Jacobians follow completely analogously. We repeated the
computations and found only three classes of Jacobians,
\begin{align}
\INP[1]_{\mc S_{\{3,4\}}} = &
-\frac{1}{16s_{12}^3}\oint\frac{dz}{z(z+\chi)}\;, \\[1.5mm]
\INP[1]_{\mc S_{\{7,8\}}} = &
-\frac{1}{16s_{12}^3}\oint\frac{dz}{z(z-\chi-1)}\;, \\[1.5mm]
\INP[1]_{\mc S_{\{1,2,5,6\}}} = &
-\frac{1}{16s_{12}^3}
\oint \frac{dz}{z(z-\chi)(z-\chi-1)}\;,
\end{align}
with composite leading singularities or simply Jacobian poles located at 
$z\in\{0,-\chi\}$, $z\in\{0,\chi+1\}$ and $z\in\{0,\chi,\chi+1\}$ respectively.
Encircling one of these global poles effectively imposes an eighth condition in
addition to the hepta-cut constraints such that the integral localizes
completely to a point in $\C^4\times \C^4$. 

Notice that the overall normalization of the Jacobians is the same for all cut
solutions and hence irrelevant in the augmented hepta-cut. In subsequent sections
we will frequently refer to integrands without the common prefactor by $J_i(z)$.

\subsection{Augmentation of Global Poles}
We realize that the product of six tree amplitudes onto which the amplitude
integrand factorizes on the hepta-cut for the present parametrization is a
holomorphic function of $z$ and therefore has no poles, except at complex 
infinity. This is in contrast to the maximal cut of the planar double box which
develops a pole at a finite value of $z$. Possible nontrivial contributions from 
poles at infinity in either of the two loop momenta are however safely ignored 
because the sum of all residues of a meromorphic function on the Riemann sphere 
must vanish identically.

\begin{figure}[!h]
\bc
\includegraphics[scale=0.65]{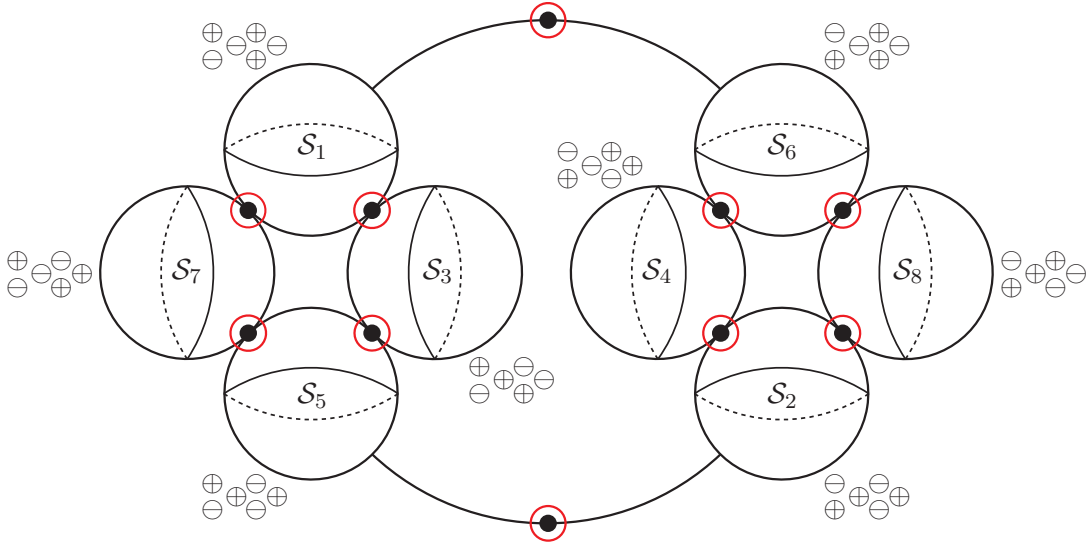}
\put(-264,152){$\mc S_1$}
\put(-300,195){$\oplus\;\;\,\oplus$}
\put(-291,190){$\ominus\;\;\hspace*{.2mm}\ominus$}
\put(-300,185){$\ominus\;\;\,\oplus$}
\put(-217,105){$\mc S_3$}
\put(-200,69){$\oplus\;\;\,\ominus$}
\put(-191,64){$\oplus\;\;\hspace*{.2mm}\ominus$}
\put(-200,59){$\ominus\;\;\,\oplus$}
\put(-264,58){$\mc S_5$}
\put(-300,25){$\oplus\;\;\,\ominus$}
\put(-291,20){$\oplus\;\;\hspace*{.2mm}\oplus$}
\put(-300,15){$\ominus\;\;\,\ominus$}
\put(-311,105){$\mc S_7$}
\put(-373,109){$\oplus\;\;\,\ominus$}
\put(-364,104){$\ominus\;\;\hspace*{.2mm}\oplus$}
\put(-373,99){$\ominus\;\;\,\oplus$}
\put(-88,152){$\mc S_6$}
\put(-67,195){$\ominus\;\;\,\oplus$}
\put(-58,190){$\ominus\;\;\hspace*{.2mm}\ominus$}
\put(-67,185){$\oplus\;\;\,\oplus$}
\put(-41,105){$\mc S_8$}
\put(-1,109){$\ominus\;\;\,\oplus$}
\put(8,104){$\oplus\;\;\hspace*{.2mm}\ominus$}
\put(-1,99){$\oplus\;\;\,\ominus$}
\put(-88,58){$\mc S_2$}
\put(-67,25){$\ominus\;\;\,\ominus$}
\put(-58,20){$\oplus\;\;\hspace*{.2mm}\oplus$}
\put(-67,15){$\oplus\;\;\,\ominus$}
\put(-135,105){$\mc S_4$}
\put(-167,150){$\ominus\;\;\,\oplus$}
\put(-158,145){$\ominus\;\;\hspace*{.2mm}\oplus$}
\put(-167,140){$\oplus\;\;\,\ominus$}
\caption{A view of the global structure of the eight on-shell solutions for the 
massless twoloop crossed box. The set of solutions has ten intersections and 
each branch is topologically equivalent to a Riemann sphere. Our convention is
to denote holomorphic and antiholomorphic vertices is by $\oplus$ and $\ominus$
respectively.}
\label{XBOXTOPOLOGY}
\ec
\end{figure}

Therefore we naively consider $4\times 3 + 4\times 2 = 20$ residues originating 
from composite leading singularities. It turns out that only some of these
contributions are in fact independent. Indeed, using several nontrivial 
relations across the on-shell branches we are able to clear out all redundancy 
and identify only ten true global residues of which the master integral 
coefficient may be built. For each relation we assume that $\xi(\ell_1,\ell_2)$ 
is holomorphic on the two Jacobian poles in question, but otherwise arbitrary. 
In our calculations, $\xi$ is of course really just a shorthand for the 
intermediate state sum of tree amplitudes on the hepta-cut. We list all
intersections of the Riemann spheres below and refer to 
fig.~\ref{XBOXTOPOLOGY} for a graphical depiction.
\begin{align}
\Res_{z=0}J_1(z)\xi(\ell_1\ell_2)\big|_{\mc S_1} = {} &
\Res_{z=0}J_6(z)\xi(\ell_1\ell_2)\big|_{\mc S_6}\; \nn \\[2mm]
\Res_{z=0}J_2(z)\xi(\ell_1\ell_2)\big|_{\mc S_2} = {} &
\Res_{z=0}J_5(z)\xi(\ell_1\ell_2)\big|_{\mc S_5}\; \nn \\[2mm]
\Res_{z=\chi}J_1(z)\xi(\ell_1\ell_2)\big|_{\mc S_1} = {} & 
\Res_{z=-\chi}J_3(z)\xi(\ell_1\ell_2)\big|_{\mc S_3}\; \nn \\[2mm]
\Res_{z=\chi}J_2(z)\xi(\ell_1\ell_2)\big|_{\mc S_2} = {} &
\Res_{z=-\chi}J_4(z)\xi(\ell_1\ell_2)\big|_{\mc S_4}\; \nn \\[2mm]
\Res_{z=\chi+1}J_1(z)\xi(\ell_1\ell_2)\big|_{\mc S_1} = {} &  
\Res_{z=0}J_7(z)\xi(\ell_1\ell_2)\big|_{\mc S_7}\; \nn \\[2mm]
\Res_{z=\chi+1}J_2(z)\xi(\ell_1\ell_2)\big|_{\mc S_2} = {} &
\Res_{z=0}J_8(z)\xi(\ell_1\ell_2)\big|_{\mc S_8}\; \nn \\[2mm]
\Res_{z=\chi+1}J_5(z)\xi(\ell_1\ell_2)\big|_{\mc S_5} = {} & 
\Res_{z=\chi+1}J_7(z)\xi(\ell_1\ell_2)\big|_{\mc S_7}\; \nn \\[2mm]
\Res_{z=\chi+1}J_6(z)\xi(\ell_1\ell_2)\big|_{\mc S_6} = {} &
\Res_{z=\chi+1}J_8(z)\xi(\ell_1\ell_2)\big|_{\mc S_8}\; \nn \\[2mm]
\Res_{z=\chi}J_5(z)\xi(\ell_1\ell_2)\big|_{\mc S_5} = {} & -
\Res_{z=0}J_3(z)\xi(\ell_1\ell_2)\big|_{\mc S_3}\; \nn \\[2mm]
\Res_{z=\chi}J_6(z)\xi(\ell_1\ell_2)\big|_{\mc S_6} = {} &  -
\Res_{z=0}J_4(z)\xi(\ell_1\ell_2)\big|_{\mc S_4}\;.
\end{align}

It is possible to use intersection labels instead,
\begin{align}
\omega_{1\cap3}\,,\;
\omega_{1\cap6}\,,\;
\omega_{1\cap7}\,,\;
\omega_{2\cap4}\,,\;
\omega_{2\cap5}\,,\;
\omega_{2\cap6}\,,\;
\omega_{3\cap5}\,,\;
\omega_{4\cap6}\,,\;
\omega_{5\cap7}\,,\;
\omega_{6\cap8}\,,
\end{align}
but contour weights with explicit reference to type of pole are more convenient
in actual calculations.

The displayed relations imply major simplications in the augmented hepta-cut 
and allow us to cut computation of residues in half. Indeed, we select only
solutions $1,2,5,6$ and avoid double counting at $z = 0$. The global poles may 
be organized using the following contour weights or generalized winding numbers,
\begin{align*}
a_{1,j} \; \longrightarrow & \;\; 
\text{encircling} \;\; z = 0 \;\; \text{for solution} \;\; \mc S_j\;, \\
a_{2,j} \; \longrightarrow & \;\; 
\text{encircling} \;\; z = \chi \;\; \text{for solution} \;\; \mc S_j\;, \\
a_{3,j} \; \longrightarrow & \;\; 
\text{encircling} \;\; z = \chi+1 \;\; \text{for solution} \;\; \mc S_j\;.
\end{align*}
We then have the following ten eight-tori encircling the global poles,
\begin{align}
T_{1,1} = {} & 
T_0\times C_{\alpha_1}(\chi)\times C_{\alpha_3}(-\chi(\chi+1))\times 
C_{\alpha_4}(0)\times C_{\beta_3=z}(0)\times C_{\beta_4}(0) \nn \\[2mm]
T_{1,2} = {} & 
T_0\times C_{\alpha_1}(\chi)\times C_{\alpha_3}(0)\times 
C_{\alpha_4}(-\chi(\chi+1))\times C_{\beta_3}(0)\times C_{\beta_4=z}(0) \nn \\[2mm]
T_{2,1} = {} & 
T_0\times C_{\alpha_1}(0)\times C_{\alpha_3}(-\chi)\times 
C_{\beta_4}(0)\times C_{\beta_3=z}(\chi)\times C_{\beta_4}(0) \nn \\[2mm]
T_{2,2} = {} & 
T_0\times C_{\alpha_1}(0)\times C_{\alpha_3}(0)\times
C_{\alpha_4}(-\chi)\times C_{\beta_3}(0)\times C_{\beta_4=z}(\chi) \nn \\[2mm]
T_{2,5} = {} & 
T_0\times C_{\alpha_1}(0)\times C_{\alpha_3}(0)\times
C_{\alpha_4}(0)\times C_{\beta_3=z}(\chi)\times C_{\beta_4}(0) \nn \\[2mm]
T_{2,6} = {} &
T_0\times C_{\alpha_1}(0)\times C_{\alpha_3}(0)\times 
C_{\alpha_4}(0)\times C_{\beta_3}(0)\times C_{\beta_4=z}(\chi) \nn \\[2mm]
T_{3,1} = {} & 
T_0\times C_{\alpha_1}(-1)\times C_{\alpha_3}(0)\times 
C_{\alpha_4}(0)\times C_{\beta_3=z}(\chi+1)\times C_{\beta_4}(0) \nn \\[2mm]
T_{3,2} = {} & 
T_0\times C_{\alpha_1}(-1)\times C_{\alpha_3}(0)\times 
C_{\alpha_4}(0)\times C_{\beta_3}(0)\times C_{\beta_4=z}(\chi+1) \nn \\[2mm]
T_{3,5} = {} & 
T_0\times C_{\alpha_1}(-1)\times 
C_{\alpha_3}(0)\times C_{\alpha_4}(\chi+1)\times 
C_{\beta_3=z}(\chi+1)\times C_{\beta_4}(0) \nn \\[2mm]
T_{3,6} = {} & 
T_0\times C_{\alpha_1}(-1)\times 
C_{\alpha_3}(\chi+1)\times C_{\alpha_4}(0)\times 
C_{\beta_3}(0)\times C_{\beta_4=z}(\chi+1)
\end{align}
where $T_0$ is the contour common to all global poles capturing parameters
which turn out to be constant on the hepta-cut,
\begin{align}
T_0 = C_{\alpha_2}(0)\times C_{\beta_1}(0)\times C_{\beta_2}(0)\;.
\end{align}

Let us now consider the localization of the master integrals realized by
expanding them onto the ten eight-tori. It turns out that, for simplicity 
say, $\INP[1]$ and $\INP[(\ell_1\cdot k_3)]$ may be chosen as master 
integrals for the crossed box topology. Evaluation of the primitive amplitude 
in this integral basis,
\begin{align}
A^{\text{2-loop}}_{\text{xbox}} = 
c_1\,\INP[1]+
c_2\,\INP[(\ell_1\cdot k_3)]+\cdots\;,
\end{align}
thus reduces the problem to determination of the rational coefficients $c_1$ 
and $c_2$ from the augmented hepta-cut. This choice of basis integrals allows 
us to directly compare our results with those of Badger, Frellesvig and 
Zhang \cite{Badger:2012dp}. All other integrals with fewer than seven 
propagators have been suppressed. In general we have
\begin{align}
2\ell_1\cdot k_3 = s_{12}\left(
-(1+\chi)\alpha_1+\chi\alpha_2-\alpha_3-\alpha_4\right)
\end{align}
and therefore,
\begin{align}
\ell_1\cdot k_3|_{\mc S_1} = 
\ell_1\cdot k_3|_{\mc S_2} = \frac{s_{12}}{2}z\;, \quad
\ell_1\cdot k_3|_{\mc S_5} = 
\ell_1\cdot k_3|_{\mc S_6} = 0\;.
\end{align}
The cut master integrals are
\begin{align}
\INP[1]_{\text{cut}} =
-\frac{1}{16s_{12}^3}
\Bigg\{
\sum_{j=1,2}
\frac{a_{1,j}}{\chi(1+\chi)}-
\sum_{j=1,2,5,6}\left(
\frac{a_{2,j}}{\chi}
-\frac{a_{3,j}}{1+\chi}\right)\Bigg\}\;,
\end{align}
\begin{align}
\INP[(\ell_1\cdot k_3)]_{\text{cut}} =
\frac{1}{32s_{12}^2}
\sum_{j=1,2}\left\{a_{2,j}-a_{3,j}\right\}\;.
\end{align}
We cancel overall factors and derive the augmented hepta-cut
\begin{align}
\sum_{i=1,2,5,6}\oint_{\Gamma_i}\frac{dz}{z(z-\chi)(z-\chi-1)}
\sum_{\substack{\text{helicities}\\\text{particles}}}
\prod_{j=1}^6 A_{(j)}^\tree(z)\big|_{\mc S_i} \hspace*{5.5cm} \nn \\ =
c_1
\,\Bigg\{
\sum_{j=1,2}
\frac{a_{1,j}}{\chi(1+\chi)}
-\sum_{j=1,2,5,6}\left(
\frac{a_{2,j}}{\chi}
-\frac{a_{3,j}}{1+\chi}\right)\Bigg\}-
\frac{s_{12}c_2}{2}
\sum_{j=1,2}\left\{a_{2,j}-a_{3,j}\right\}\;.
\end{align}
The intermediate state sum over the product of six tree amplitudes takes the
explicit form
\begin{align}
\;
\sum_{\substack{\text{particles}\\ \text{helicities}}}
\prod_{j=1}^6 A_{(j)}^\tree(z)\big|_{\mc S_i} \hspace*{11cm} \nn \\ 
= \sum_{\text{particles}} \sum_{\lambda_i = \pm}
A_{(1)}^\tree(-p_1^{-\lambda_1},k_1,p_2^{\lambda_3})
A_{(2)}^\tree(-p_6^{-\lambda_6},k_2,p_7^{\lambda_7})
A_{(3)}^\tree(-p_3^{-\lambda_2},k_3,p_4^{\lambda_4}) 
\nn \hspace*{1.25cm} \\ \times\,
A_{(4)}^\tree(-p_4^{-\lambda_4},k_4,p_5^{\lambda_5})
A_{(5)}^\tree(-p_5^{-\lambda_5},p_1^{\lambda_1},p_6^{\lambda_6})
A_{(6)}^\tree(-p_2^{-\lambda_2},p_3^{\lambda_3},-p_7^{-\lambda_7})
\Big|_{\mc S_i}\;,
\vspace*{.2cm}
\end{align}
where, in this notation, $p_i$ is the $i$th inverse propagator of the crossed 
box diagram, obtained by following momentum flow with the initial 
identification $p_1 = \ell_1+k_1$.

\subsection{Integral Reduction Identities}
In order to constrain the integration contours we impose consistency conditions. 
It is completely clear that vanishing Feynman integrals should have vanishing
hepta-cuts. Otherwise the unitarity procedure is not well-defined. Equivalently,
we can demand that any integral identity is preserved,
\begin{align}
I_1 = I_2 \LRa I_{1,\text{cut}} = I_{2,\text{cut}}\;.
\end{align}

We identify the complete variety of Levi-Civita symbols that appears in integral 
reduction, after using momentum conservation, and require continued vanishing of 
the following five integrals after pushing loop integration from real slices 
into $\C^4\times \C^4$,
\begin{gather}
\INP[\varepsilon(\ell_1,k_2,k_3,k_4)]\;, \quad
\INP[\varepsilon(\ell_2,k_2,k_3,k_4)]\;, \nn \\
\INP[\varepsilon(\ell_1,\ell_2,k_1,k_2)]\;, \quad
\INP[\varepsilon(\ell_1,\ell_2,k_1,k_3)]\;, \quad
\INP[\varepsilon(\ell_1,\ell_2,k_2,k_3)]\;.
\end{gather}
Let us set the stage and evaluate the first two constraints explicitly. We expand 
the integral onto the augmented hepta-cut,
\begin{align}
0 = {} & \INP[\varepsilon(\ell_1,k_2,k_3,k_4)]_{\text{cut}}
\LLra \nn \\
0 = {} & \oint_{\Gamma_1}dz\frac{
\varepsilon\left(
(\chi-z)k_1^\mu+\frac{
s_{12}\chi(z-\chi-1)}{2\avg{14}\Avg{42}}\expval{1^-}{\gamma^\mu}{2^-}
,k_2,k_3,k_4\right)}
{z(z-\chi)(z-\chi-1)} \nn \\ 
+ {} & \oint_{\Gamma_2}dz\frac{
\varepsilon\left(
(\chi-z)k_1^\mu+\frac{
s_{12}\chi(z-\chi-1)}{2\avg{24}\Avg{41}}\expval{2^-}{\gamma^\mu}{1^-}
,k_2,k_3,k_4\right)}
{z(z-\chi)(z-\chi-1)} \nn \\
+ {} & \oint_{\Gamma_5}dz\frac{
\varepsilon\left(
(\chi-z)k_1^\mu+\frac{
s_{12}(\chi+1)(z-\chi)}{2\avg{24}\Avg{41}}\expval{2^-}{\gamma^\mu}{1^-}
,k_2,k_3,k_4\right)}
{z(z-\chi)(z-\chi-1)} \nn \\
+ {} & \oint_{\Gamma_6}dz\frac{
\varepsilon\left(
(\chi-z)k_1^\mu+\frac{
s_{12}(\chi+1)(z-\chi)}{2\avg{14}\Avg{42}}\expval{1^-}{\gamma^\mu}{2^-}
,k_2,k_3,k_4\right)}
{z(z-\chi)(z-\chi-1)}
\end{align}
and in virtue of the relation
\begin{align}
\varepsilon\left(
\frac{\expval{1^-}{\gamma^\mu}{2^-}}
{\avg{14}\Avg{42}},k_2,k_3,k_4\right) = -
\varepsilon\left(
\frac{\expval{2^-}{\gamma^\mu}{1^-}}
{\avg{24}\Avg{41}},k_2,k_3,k_4\right)
\end{align}
we then obtain the constraint equation,
\begin{align}
0 = \INP[\varepsilon(\ell_1,k_2,k_3,k_4)]_{\text{cut}} =
a_{1,1}-a_{1,2}-a_{2,1}+a_{2,2}+a_{3,5}-a_{3,6} = 0\;.
\end{align}
Likewise, the second vanishing identity in question,
\begin{align}
0 = {} & \INP[\varepsilon(\ell_2,k_2,k_3,k_4)]_{\text{cut}} 
\LLra \nn \\
0 = {} & \oint_{\Gamma_1+\Gamma_5}dz\frac{
\varepsilon\left(
\frac{s_{12}z}{2\avg{31}\Avg{14}}\expval{3^-}{\gamma^\mu}{4^-}
,k_2,k_3,k_4\right)}
{z(z-\chi)(z-\chi-1)} \nn \\ 
+ {} & \oint_{\Gamma_2+\Gamma_6}dz\frac{
\varepsilon\left(
\frac{s_{12}z}{2\avg{41}\Avg{13}}\expval{4^-}{\gamma^\mu}{3^-}
,k_2,k_3,k_4\right)}
{z(z-\chi)(z-\chi-1)}\;, 
\end{align}
linearity in the contour subscript being implied, becomes
\begin{align}
0 = \INP[\varepsilon(\ell_2,k_2,k_3,k_4)]_{\text{cut}} =
a_{2,1}-a_{2,2}-a_{3,1}+a_{3,2}+a_{2,5}-a_{2,6}-a_{3,5}+a_{3,6}\;,
\end{align}
where we used the fact that
\begin{align}
\varepsilon\left(
\frac{\expval{3^-}{\gamma^\mu}{4^-}}
{\avg{31}\Avg{14}},k_2,k_3,k_4\right) = -
\varepsilon\left(
\frac{\expval{4^-}{\gamma^\mu}{3^-}}
{\avg{41}\Avg{13}},k_2,k_3,k_4\right)\;.
\end{align}
The last three parity vanishing requirements,
\begin{align}
0 = {} & \INP[\varepsilon(\ell_1,\ell_2,k_i,k_j)] \LLra \nn \\
0 = {} & \oint_{\Gamma_1}dz\frac{\varepsilon\left(
(\chi-z)k_1^\mu+\frac{s_{12}\chi(z-\chi-1)}{2\avg{14}\Avg{42}}
\expval{1^-}{\gamma^\mu}{2^-},
\frac{s_{12}z}{2\avg{31}\Avg{14}}\expval{3^-}{\gamma^\mu}{4^-},
k_i,k_j\right)}
{z(z-\chi)(z-\chi-1)} \nn \\
+ {} & \oint_{\Gamma_2}dz\frac{\varepsilon\left(
(\chi-z)k_1^\mu+\frac{s_{12}\chi(z-\chi-1)}{2\avg{24}\Avg{41}}
\expval{2^-}{\gamma^\mu}{1^-},
\frac{s_{12}z}{2\avg{41}\Avg{11}}\expval{4^-}{\gamma^\mu}{3^-},
k_i,k_j\right)}
{z(z-\chi)(z-\chi-1)} \nn \\
+ {} & \oint_{\Gamma_5}dz\frac{\varepsilon\left(
(\chi-z)k_1^\mu+\frac{s_{12}(\chi+1)(z-\chi)}{2\avg{24}\Avg{41}}
\expval{2^-}{\gamma^\mu}{1^-},
\frac{s_{12}z}{2\avg{31}\Avg{14}}\expval{3^-}{\gamma^\mu}{4^-},
k_i,k_j\right)}
{z(z-\chi)(z-\chi-1)} \nn \\
+ {} & \oint_{\Gamma_6}dz\frac{\varepsilon\left(
(\chi-z)k_1^\mu+\frac{s_{12}(\chi+1)(z-\chi)}{2\avg{14}\Avg{42}}
\expval{1^-}{\gamma^\mu}{2^-},
\frac{s_{12}z}{2\avg{41}\Avg{13}}\expval{4^-}{\gamma^\mu}{3^-},
k_i,k_j\right)}
{z(z-\chi)(z-\chi-1)}
\end{align}
for $(i,j)\in\{(1,2),(1,3),(2,3)\}$ are also rather straightforward to obtain
by this strategy, so we will spare the reader for details and just quote the
final expressions,
\begin{align}
0 = \INP[\varepsilon(\ell_1,\ell_2,k_1,k_2)]_{\text{cut}} = 
a_{2,1}-a_{2,2}-a_{3,5}+a_{3,6} = 0\;, \nn \\[1mm]
0 = \INP[\varepsilon(\ell_1,\ell_2,k_1,k_3)]_{\text{cut}} =
a_{2,1}-a_{2,2} = 0\;, \nn \\[1mm]
0 = \INP[\varepsilon(\ell_1,\ell_2,k_2,k_3)]_{\text{cut}} =
a_{3,1}-a_{3,2} = 0\;.
\end{align}
Reduction together with the two single-momentum parity constraints produces the 
following five linearly independent parity vanishing identities,
\begin{align}
a_{1,1}-a_{1,2} = 0\;, \nn \\
a_{2,1}-a_{2,2} = 0\;, \nn \\ 
a_{2,5}-a_{2,6} = 0\;, \nn \\ 
a_{3,1}-a_{3,2} = 0\;, \nn \\
a_{3,5}-a_{3,6} = 0\;.
\end{align}
The displayed equations have a very simple interpretation; they simply translate 
into the statement that all contours across parity-conjugate solutions 
$\mc S_1~\longleftrightarrow~\mc S_2$ and 
$\mc S_5~\longleftrightarrow~\mc S_6$ must carry weights of equal
values, thereby resembling previous observations for both the one-loop box and the 
planar double box. Actually this feature is expected as its origin can be
traced back to the equality of the Jacobians that arise upon localization of the
crossed box integral onto the Riemann spheres parametrized by the four hepta-cut 
branches in consideration.

We next consider contour constraint equations arising from integration-by-parts 
identities used for reduction onto master integrals. There are two nonspurious 
irreducible scalar products parametrizing the general integrand. Gram matrix 
relations for four-dimensional momenta remove dependent terms and imply that we 
have the following nineteen naively irreducible tensor integrals in 
renormalizable theories,
\begin{gather}
\INP[1]\,,\; 
\mc \INP[(\ell_1\cdot k_3)]\,,\;
\INP[(\ell_1\cdot k_3)^2]\,,\;
\INP[(\ell_1\cdot k_3)^3]\,,\;
\INP[(\ell_1\cdot k_3)^4]\,,\;
\nn \\[1mm]
\INP[(\ell_2\cdot k_2)]\,,\;
\INP[(\ell_2\cdot k_2)^2]\,,\;
\INP[(\ell_2\cdot k_2)^3]\,,\;
\INP[(\ell_2\cdot k_2)^4]\,,\;
\nn \\[1mm]
\INP[(\ell_2\cdot k_2)^5]\,,\;
\INP[(\ell_2\cdot k_2)^6]\,,\;
\INP[(\ell_1\cdot k_3)(\ell_2\cdot k_2)]\,,\;
\INP[(\ell_1\cdot k_3)^2(\ell_2\cdot k_2)]\,,\;
\nn \\[1mm]
\INP[(\ell_1\cdot k_3)^3(\ell_2\cdot k_2)]\,,\;
\INP[(\ell_1\cdot k_3)^4(\ell_2\cdot k_2)]\,,\;
\INP[(\ell_1\cdot k_3)(\ell_2\cdot k_2)^2]\,,\;
\nn \\[1mm]
\INP[(\ell_1\cdot k_3)^2(\ell_2\cdot k_2)^2]\,,\;
\INP[(\ell_1\cdot k_3)^3(\ell_2\cdot k_2)^2]\,,\;
\INP[(\ell_1\cdot k_3)^4(\ell_2\cdot k_2)^2]\,.
\end{gather}
All identities can be generated with the Mathematica package {\tt FIRE} and are 
listed in appendix~\ref{IBPXBOX}. It now just remains to evaluate all tensor 
integrals on the augmented hepta-cut and enforce continued validity of the 
integral reduction equations. To this end we compute the tensors using the 
parametrized loop momenta, 
\begin{align}
\ell_2\cdot k_2 = \frac{s_{12}}{2}\left(
\chi\beta_1-(1+\chi)\beta_2-\beta_3-\beta_4\right)\;,
\end{align}
such that on the relevant on-shell branches,
\begin{align}
\ell_2\cdot k_2|_{\mc S_1} =
\ell_2\cdot k_2|_{\mc S_2} =
\ell_2\cdot k_2|_{\mc S_5} =
\ell_2\cdot k_2|_{\mc S_6} = -\frac{s_{12}}{2}z\;.
\end{align}
Then we can write down the augmented hepta-cuts
\begin{align}
\INP[(\ell_1\cdot k_3)^n]_{\text{cut}} = {} &
-\frac{1}{16s_{12}^3}\left(\frac{s_{12}}{2}\right)^n
\sum_{i=1,2}\oint_{\Gamma_i}dz\frac{z^{n-1}}{(z-\chi)(z-\chi-1)}\;, \nn \\[3mm]
\INP[(\ell_2\cdot k_2)^m]_{\text{cut}} = {} &
\frac{(-1)^{m+1}}{16s_{12}^3}\left(\frac{s_{12}}{2}\right)^m
\sum_{i=1,2,5,6}\oint_{\Gamma_i}dz\frac{z^{m-1}}{(z-\chi)(z-\chi-1)}\;, \nn
\\[3mm]
\INP[(\ell_1\cdot k_3)^n(\ell_2\cdot k_2)^m]_{\text{cut}} = {} &
\frac{(-1)^{m+1}}{16s_{12}^3}\left(\frac{s_{12}}{2}\right)^{n+m}
\sum_{i=1,2}\oint_{\Gamma_i} dz\frac{z^{n+m-1}}{(z-\chi)(z-\chi-1)}\;,
\end{align}
and obtain the explicit relations
\begin{align}
\INP[(\ell_1\cdot k_3)^n]_{\text{cut}} = {} & 
\frac{1}{16s_{12}^3}\left(\frac{s_{12}}{2}\right)^n
\sum_{j=1,2}\left\{\chi^{n-1}a_{2,j}-(1+\chi)^{n-1}a_{3,j}\right\}\;, \nn
\\[3mm]
\INP[(\ell_2\cdot k_2)^m)]_{\text{cut}} = {} &
\frac{(-1)^m}{16s_{12}^3}\left(\frac{s_{12}}{2}\right)^m
\sum_{j=1,2,5,6}\left\{\chi^{m-1}a_{2,j}-(1+\chi)^{m-1}a_{3,j}\right\}\;, \nn
\\[3mm]
\INP[(\ell_1\cdot k_3)^n(\ell_2\cdot k_2)^m]_{\text{cut}} = {} &
\frac{(-1)^m}{16s_{12}^3}\left(\frac{s_{12}}{2}\right)^{n+m}
\sum_{j=1,2}\left\{\chi^{n+m-1}a_{2,j}-(1+\chi)^{n+m-1}a_{3,j}\right\}\;.
\end{align}
Insertion into the integration by parts identities yields seventeen linear
relations among the winding numbers. We are able able to clarify any redundancy 
and derive only three independent constraints,
\begin{align}
a_{2,1}+a_{2,2}-a_{2,5}-a_{2,6} = 0\;, \nn \\
a_{1,1}+a_{1,2}+a_{2,1}+a_{2,2}+a_{3,1}+a_{3,2} = 0\;, \nn \\
a_{1,1}+a_{1,2}+a_{2,1}+a_{2,2}+a_{3,5}+a_{3,6} = 0\;.
\end{align}
We further compress these equations together with the parity vanishing 
identities and find the final form of the eight constraint equations,
\begin{align}
a_{1,1}-a_{1,2} = a_{2,1}-a_{2,2} = a_{2,5}-a_{2,6} =
a_{3,1}-a_{3,2} = a_{3,5}-a_{3,6} = 0\;, \nn \\
a_{2,1}-a_{2,5} = a_{3,1}-a_{3,5} = 0\;, \nn \\
a_{1,1}+a_{2,1}+a_{3,1} = 0\;.
\end{align}
In addition to the requirements arising from Levi-Civita integrals we see that
winding numbers of each type of global pole must be uniform across all on-shell
solutions, whereas the last equation states that the sum of weights within a
branch vanishes.

\subsection{Unique Master Integral Projectors}
We fix the remaining freedom of the contour weights and derive independent master 
contours that each project out a single master integral coefficient, for instance 
we isolate the scalar master integral by imposing the conditions,
\begin{align}
\sum_{j=1,2}\left\{a_{2,j}-a_{3,j}\right\} = 0\;, \quad
\sum_{j=1,2}
\frac{a_{1,j}}{\chi(1+\chi)}
-\sum_{j=1,2,5,6}\left(
\frac{a_{2,j}}{\chi}
-\frac{a_{3,j}}{1+\chi}\right) = 1\;.
\end{align}
and vice versa for the tensor master integral,
\begin{align}
\INP[1]_{\text{cut}} = 0\;, \quad
\INP[(\ell_1\cdot k_3)]_{\text{cut}} = -\frac{2}{s_{12}}\;.
\end{align}
The displayed normalization conditions are chosen purely for convenience in order 
for the contour weights to soak up overall prefactors. The cost is loss of an
immediate geometrical interpretation of the contour weights as integral winding 
numbers. We solve the two set of equations and find two master contours which we 
denote $\mc M_1$ and $\mc M_2$ respectively.
\begin{align}
\begin{array}{cc}
{} & a_{1,1} = a_{1,2} = \frac{1}{4}\chi(1+\chi) \\[2mm]
\mc M_1\,: \quad\quad & a_{2,1} = a_{2,2} = a_{2,5} = a_{2,6} =
-\frac{1}{8}\chi(1+\chi) \\[2mm]
{} & a_{3,1} = a_{3,2} = a_{3,5} = a_{3,6} = -\frac{1}{8}\chi(1+\chi) \\[5mm]
{} & a_{1,1} = a_{1,2} = -\dfrac{1+2\chi}{2s_{12}} \\[3mm]
\mc M_2\,: \quad\quad & a_{2,1} = a_{2,2} = a_{2,5} = a_{2,6} =
-\dfrac{1-2\chi}{4s_{12}} \\[4mm]
{} & a_{3,1} = a_{3,2} = a_{3,5} = a_{3,6} = \dfrac{3+2\chi}{4s_{12}}
\end{array}
\end{align}
Therefore our final formula for the master integral coefficients can be written
in the remarkably compact form
\begin{align}
\boxed{
c_i = 
\oint_{\mc M_i}
\frac{dz}{z(z-\chi)(z-\chi-1)}
\sum_{\substack{\text{helicities}\\\text{particles}}}
\prod_{j=1}^6 A_{(j)}^\tree(z)\;,
}
\end{align}
or as explicitly as expansions in residues,
\begin{align}
c_1 = {} &
\frac{1}{4}
\sum_{i = 1,2}\Res_{z=0}\frac{1}{z}
\sum_{\substack{\text{particles}\\ \text{helicities}}}
\prod_{j=1}^6 A_{(j)}^\tree(z)\big|_{\mc S_i} \nn \\ &
+\frac{1+\chi}{8}
\sum_{i = 1,2,5,6}\Res_{z=\chi}\frac{1}{z-\chi}
\sum_{\substack{\text{particles}\\ \text{helicities}}}
\prod_{j=1}^6 A_{(j)}^\tree(z)\big|_{\mc S_i} \nn \\ &
-\frac{\chi}{8}
\sum_{i = 1,2,5,6}\Res_{z=\chi+1}\frac{1}{z-\chi-1}
\sum_{\substack{\text{particles}\\ \text{helicities}}}
\prod_{j=1}^6 A_{(j)}^\tree(z)\big|_{\mc S_i}\;, \\[3mm]
c_2 = {} &
-\frac{1+2\chi}{2s_{12}\chi(\chi+1)}
\sum_{i = 1,2}\Res_{z=0}\frac{1}{z}
\sum_{\substack{\text{particles}\\ \text{helicities}}}
\prod_{j=1}^6 A_{(j)}^\tree(z)\big|_{\mc S_i} \nn \\ &
+\frac{1-2\chi}{4s_{12}\chi}
\sum_{i = 1,2,5,6}\Res_{z=\chi}\frac{1}{z-\chi}
\sum_{\substack{\text{particles}\\ \text{helicities}}}
\prod_{j=1}^6 A_{(j)}^\tree(z)\big|_{\mc S_i} \nn \\ &
+\frac{3+2\chi}{4s_{12}(\chi+1)}
\sum_{i = 1,2,5,6}\Res_{z=\chi+1}\frac{1}{z-\chi-1}
\sum_{\substack{\text{particles}\\ \text{helicities}}}
\prod_{j=1}^6 A_{(j)}^\tree(z)\big|_{\mc S_i}\;.
\label{NONPLANARFORMULAE}
\end{align}
Although the latter expressions at first sight may look slightly complicated, 
notice that the number of ingredients really is minimal. Indeed, once the 
intermediate state sum is computed on the four on-shell branches, which is 
rather elementary, it is just a matter of plugging in values of $z$ appropriate 
to the residues and forming the indicated linear combinations to get both 
master integral coefficients. 

We finally remark that the formulae in this paper are of course compatible with 
the Bern-Carrasco-Johansson (BCJ) color/kinematics duality 
\cite{Bern:2008qj,Bern:2010ue} in the maximally supersymmetric case. Indeed, by 
absence of triangle subgraphs in $\mc N = 4$ we expect the master integral 
coefficients for the planar and nonplanar double boxes to be equal. The 
intermediate state sum in $\mc N = 4$ Yang-Mills theory is independent of both 
loop momenta and a standard result in the litterature. Anyway, it is easy to 
rederive for both topologies,
\begin{align}
\sum_{\substack{\mc N = 4\\ \text{multiplet}}}
\prod_{j=1}^6 A_{(j)}^\tree(z)\big|_{\mc S_i} = -s_{12}^2s_{14}A^\tree_4\;.
\end{align}
Then we readily get
$c_{1;\mc N = 4}^{\mathrm{dbox}} = c_{1;\mc N = 4}^{\mathrm{xbox}} = 
-s_{12}^2s_{14}A^\tree_4$ and 
$c_{2,\mc N = 4}^{\mathrm{dbox}} = c_{2,\mc N = 4}^{\mathrm{xbox}} = 0$.

\section{Examples}
In this section we apply the master integral formulae to two-loop four-point 
gluon amplitudes with specific helicity configurations. We only consider
hepta-cuts in the $s$-channel, because contributions from the $t$-channel can be
obtained completely analogously. To account for the cyclic permutation we should
however substitute $\chi\to\chi^{-1}$ in \eqref{NONPLANARFORMULAE}. Our results 
are valid for supersymmetric theories with $\mc N$ supersymmetries including 
QCD. 

We track contributions to the intermediate state sums using superspace
techniques developed in \cite{Bern:2009xq,Sogaard:2011pr}. In particular, we 
exploit that the transition from $\mc N = 4$ to fewer supersymmetries is very 
straightforward,   
\begin{align}
\sum_{\substack{\mc N = 4\\ \text{multiplet}}}
\prod_{i=1}^k A_{(i)}^\tree = 
\Delta^{-1}(A+B+C+\cdots)^4 \longrightarrow \hspace*{6.3cm} \nn \\
\sum_{\substack{\mc N < 4 \\ \text{multiplet}}}
\prod_{i=1}^k A_{(i)}^\tree = 
\Delta^{-1}(A+B+C+\cdots)^{\mc N}(A^{4-\mc N}+B^{4-\mc N}+C^{4-\mc N}+\cdots)\;.
\end{align}
Here $A,B,C,\dots$ contain spin factors for each kinematically valid assignment 
of helicities on the internal lines with only gluons propagating the in loops 
whereas $\Delta$ is the denominator of the supersum. Let us consider the case of 
only two gluonic contributions $A$ and $B$ in more detail. This situation is 
relevant for quadruple cuts of one-loop amplitudes and hepta-cuts at two loops 
for instance. The trick is to expand the state sum around $A = -B$ such that 
\cite{Kosower:2011ty}
\begin{align}
\sum_{\substack{\mc N\leq 4\\ \text{multiplet}}}
\prod_{j=1}^6 A_{(j)}^\tree = {} &
\frac{A^{4-\mc N}+B^{4-\mc N}}{(A+B)^{4-\mc N}}
(1-\tfrac{1}{2}\delta_{\mc N,4})
\sum_{\substack{\mc N = 4\\ \text{multiplet}}}
\prod_{j=1}^6 A_{(j)}^\tree
\nn \\ = {} &
\bigg\{
1-(4-\mc N)\left(\frac{A}{A+B}\right)+
(4-\mc N)\left(\frac{A}{A+B}\right)^2\bigg\}
\sum_{\substack{\mc N = 4\\ \text{multiplet}}}
\prod_{j=1}^6 A_{(j)}^\tree\;.
\end{align}

In all computations we use the three- and four-gluon MHV amplitudes  
\begin{align}
A^\tree_{--+} = i\frac{\avg{12}^4}{\avg{12}\avg{23}\avg{31}}\;, \quad
A^\tree_{--++} = i\frac{\avg{12}^4}{\avg{12}\avg{23}\avg{34}\avg{41}}\;, \quad
A^\tree_{-+-+} = i\frac{\avg{13}^4}{\avg{12}\avg{23}\avg{34}\avg{41}}\;,
\end{align}
together with their parity conjugates obtained by $\avg{\,}\to\Avg{\;}$. 

\subsection{Helicities $--++$}
Our starting point is the tree-level data
\begin{align}
\sum_{\substack{\mc N = 4\\ \text{multiplet}}}
\prod_{j=1}^6 A_{(j)}^\tree(z)\big|_{\mc S_i} = -s_{12}^2s_{14}A^\tree_{--++}\;,
\end{align}
which is independent of the loop momenta.

\begin{figure}[!h]
\bc
\includegraphics[scale=0.6]{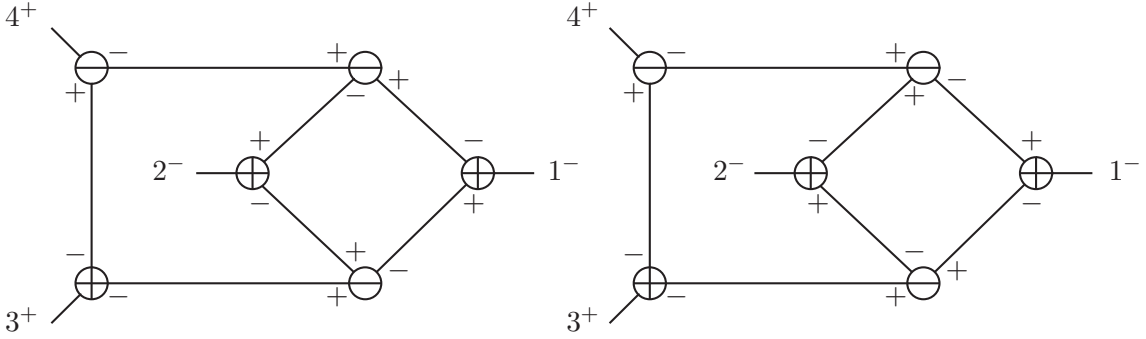} \hspace*{0.5cm}
\put(-17,57){$1^-$}
\put(-165,57){$2^-$}
\put(-220,0){$3^+$}
\put(-220,115){$4^+$}
\put(-198,87){$+$}
\put(-182,102){$-$}
\put(-198,26){$-$}
\put(-182,10){$-$}
\put(-100,10){$+$}
\put(-100,102){$+$}
\put(-77,20){$-$}
\put(-77,92){$+$}
\put(-129,69){$+$}
\put(-129,45){$-$}
\put(-49,69){$-$}
\put(-49,45){$+$}
\put(-93,86){$-$}
\put(-93,27){$+$}
\includegraphics[scale=0.6]{holo2} 
\put(2,57){$1^-$}
\put(-146,57){$2^-$}
\put(-201,0){$3^+$}
\put(-201,115){$4^+$}
\put(-180,87){$+$}
\put(-164,102){$-$}
\put(-180,26){$-$}
\put(-164,10){$-$}
\put(-82,10){$+$}
\put(-82,102){$+$}
\put(-59,20){$+$}
\put(-59,92){$-$}
\put(-111,69){$-$}
\put(-111,45){$+$}
\put(-31,69){$+$}
\put(-31,45){$-$}
\put(-75,86){$+$}
\put(-75,27){$-$}
\caption{Hepta-cut solution $\mc S_2$ allows two distinct assignments of
helicities on the internal lines in the $--++$ two-loop crossed box.}
\label{--++S1}
\ec
\end{figure}

We then compute the ratio of a general state sum relative to that of $\mc N = 4$ 
explicitly for solution $\mc S_2$ as an example. The two valid distributions of 
internal helicities denoted $A$ and $B$ are shown in fig.~\ref{--++S1} and the
depiction of holomorphic and antiholomorphic vertices by $\oplus$ and $\ominus$ 
follows \cite{Bern:2009xq}. 

The relative sign between gluonic contributions is in general specified by 
signatures of Grassmann variables in on-shell superspace and by carefully 
working out directions of all internal momenta and applying analytic 
continuations appropriately, i.e. $p_i\to-p_i$ implies change of sign for the 
holomorphic spinor while the conjugate is left unchanged. However, for our 
purposes it is advantageous to cut the calculation short and just infer the 
sign by matching the expression in $\mc N = 4$ theory, i.e. insisting that
\begin{align}
\Delta^{-1}(A+B)^4 = -s_{12}^2s_{14}A^\tree_{--++}\;.
\end{align}

To proceed, label propagators consecutively from $p_1 = \ell_1+k_1$ according
to the momentum flow previously outlined in fig.~\ref{XBOXDIAGRAM}. For
instance, $\ell_1 = p_2$ and $\ell_2 = p_4$. Then spinor strings for helicity 
configurations $A$ and $B$ are
\begin{align}
A = {} & 
\avg{2p_7}\Avg{p_7p_3}\avg{p_3p_4}\Avg{p_44}\avg{1p_1}\Avg{p_1p_5}\;, \\[1mm]
B = {} & 
-\avg{1p_2}\Avg{p_2p_3}\avg{p_3p_4}\Avg{p_44}\avg{2p_6}\Avg{p_6p_5}\;,
\end{align}
whereas the denominator reads
\begin{align}
\label{S2CUTDEN}
\Delta = {} &
\avg{p_11}\avg{1p_2}\avg{p_2p_1}
\Avg{p_2p_3}\Avg{p_3p_7}\Avg{p_7p_2}
\avg{p_33}\avg{3p_4}\avg{p_4p_3} \nn \\ {} & \times
\Avg{p_44}\Avg{4p_5}\Avg{p_5p_4}
\Avg{p_5p_1}\Avg{p_1p_6}\Avg{p_6p_5}
\avg{2p_6}\avg{p_6p_7}\avg{p_72}\;.
\end{align}
We now use momentum conservation several times to cancel common factors and 
get the compact expression
\begin{align}
\frac{A}{A+B} =
-\frac{\aAvg{2}{p_2}{2}}{s_{12}} = z-\chi\;,
\end{align}
where the last equality follows by inserting explicit values for the internal
momenta on the hepta-cut branch in question,
\begin{align}
p_1^\mu = \ell_1^\mu+k_1^\mu = -(z-\chi-1)k_1^\mu+
\frac{s_{12}z}{2\avg{24}\Avg{41}}\expval{2^-}{\gamma^\mu}{1^-}
\end{align}
and $p_2^\mu = p_1^\mu-k_1^\mu$. The state sum in case of $\mc N\leq 4$ 
supersymmetries can thus be written as
\begin{align}
\sum_{\substack{\mc N\leq 4\\ \text{multiplet}}}
\prod_{j=1}^6 A_{(j)}^\tree(z)\big|_{\mc S_2} = 
-s_{12}^2s_{14}A^\tree_{--++}\left\{1-(4-\mc N)(z-\chi)+
(4-\mc N)(z-\chi)^2\right\}\;.
\end{align}
\begin{figure}[!h]
\bc
\includegraphics[scale=0.6]{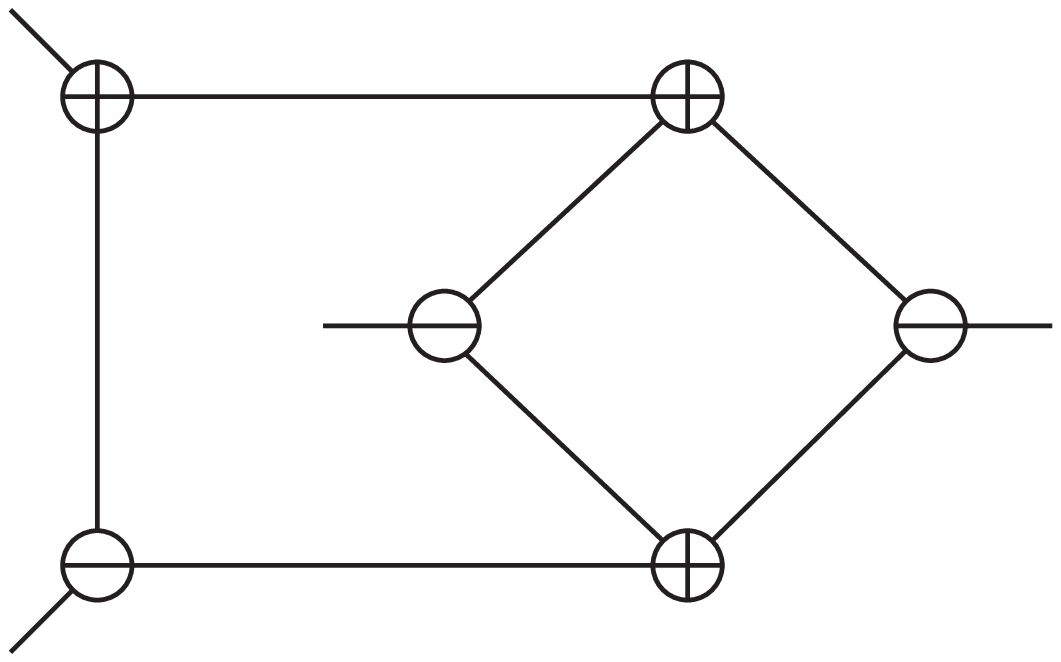} \hspace*{0.5cm}
\put(-17,57){$1^-$}
\put(-165,57){$2^-$}
\put(-220,0){$3^+$}
\put(-220,115){$4^+$}
\put(-198,87){$-$}
\put(-182,102){$-$}
\put(-198,26){$+$}
\put(-182,10){$-$}
\put(-100,10){$+$}
\put(-100,102){$+$}
\put(-77,20){$-$}
\put(-77,92){$-$}
\put(-129,69){$+$}
\put(-129,45){$+$}
\put(-49,69){$+$}
\put(-49,45){$+$}
\put(-93,86){$-$}
\put(-93,27){$-$}
\includegraphics[scale=0.6]{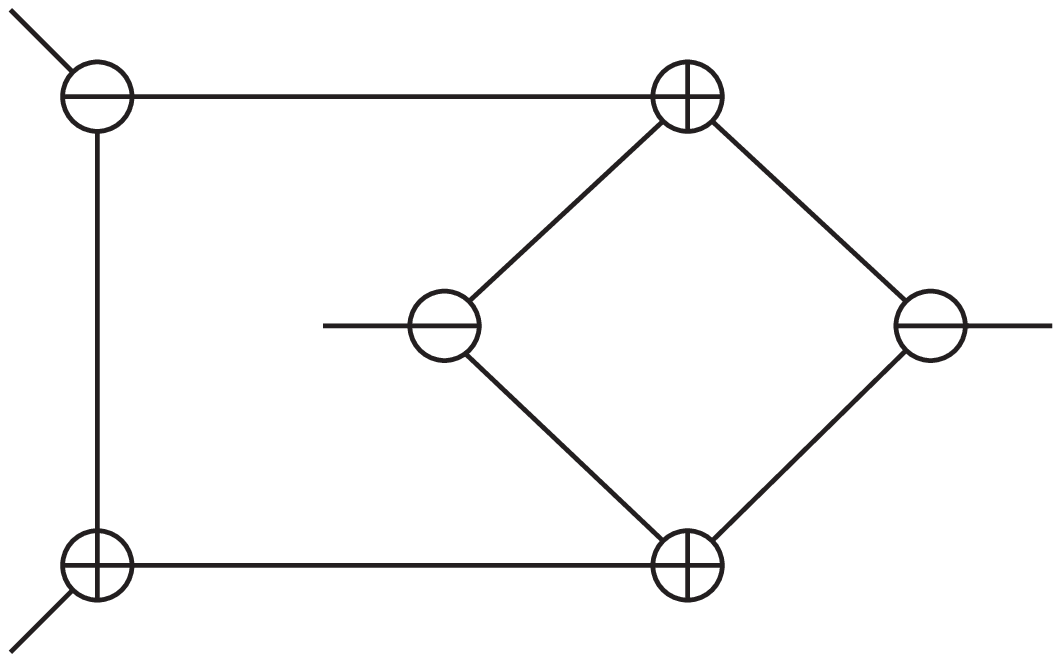} 
\put(2,57){$1^-$}
\put(-146,57){$2^-$}
\put(-201,0){$3^+$}
\put(-201,115){$4^+$}
\put(-180,87){$+$}
\put(-164,102){$-$}
\put(-180,26){$-$}
\put(-164,10){$-$}
\put(-82,10){$+$}
\put(-82,102){$+$}
\put(-59,20){$-$}
\put(-59,92){$-$}
\put(-111,69){$+$}
\put(-111,45){$+$}
\put(-31,69){$+$}
\put(-31,45){$+$}
\put(-75,86){$-$}
\put(-75,27){$-$}
\caption{Hepta-cut solutions $\mc S_1$ and $\mc S_6$ are both singlets for
external helicities $--++$ in the sense that only gluons are allowed to
propagate in the loops, thereby producing state sums that are independent of
the number of supersymmetries.}
\label{--++S1S6}
\ec
\end{figure}

The treatment is similar for the other on-shell branches. Examples of supported
helicity configurations are shown in fig.~\ref{--++S1S6}. In the end, inserting 
the multiplet sums into \eqref{NONPLANARFORMULAE} and computing all residues 
yield the master integral coefficients reconstructed to $\mc O(\epsilon^0)$ in 
$\mc N = 4,2,1,0$ Yang-Mills theory, with the result 
\begin{align}
A^{\text{xbox}}_{--++} = -s_{12}^2s_{14}A^\tree_{--++}\bigg\{
\left[1+(4-\mc N)\frac{s_{14}}{4s_{12}}
\left(1+\frac{s_{14}}{s_{12}}\right)\right]\INP[1] \nn \hspace*{3.5cm} \\[1mm]
+\,(4-\mc N)\frac{s_{13}-s_{14}}{2s_{12}^2}\INP[(\ell_1\cdot k_3)]\bigg\}\;.
\end{align}

\subsection{Helicities $-+-+$}
\begin{figure}[!h]
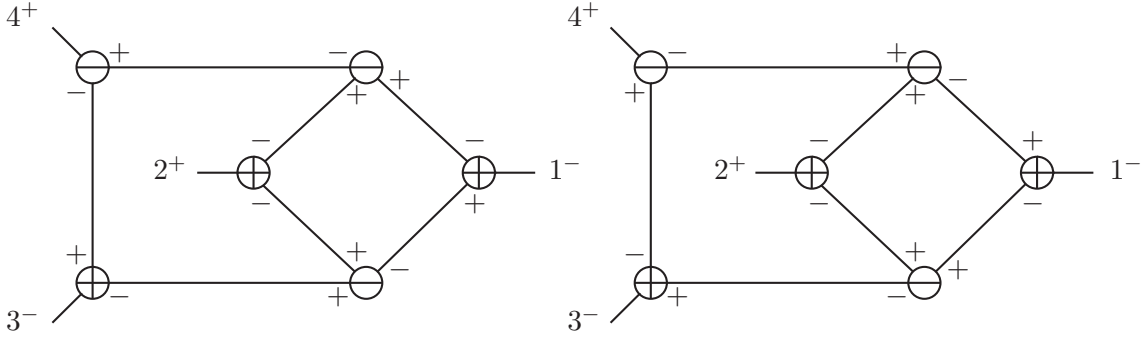

\bc
\includegraphics[scale=0.6]{holo2} \hspace*{0.5cm}
\put(-17,57){$1^-$}
\put(-165,57){$2^+$}
\put(-220,0){$3^-$}
\put(-220,115){$4^+$}
\put(-198,87){$-$}
\put(-182,102){$+$}
\put(-198,26){$+$}
\put(-182,10){$-$}
\put(-100,10){$+$}
\put(-100,102){$-$}
\put(-77,20){$-$}
\put(-77,92){$+$}
\put(-129,69){$-$}
\put(-129,45){$-$}
\put(-49,69){$-$}
\put(-49,45){$+$}
\put(-93,86){$+$}
\put(-93,27){$+$}
\includegraphics[scale=0.6]{holo2} 
\put(2,57){$1^-$}
\put(-146,57){$2^+$}
\put(-201,0){$3^-$}
\put(-201,115){$4^+$}
\put(-180,87){$+$}
\put(-164,102){$-$}
\put(-180,26){$-$}
\put(-164,10){$+$}
\put(-82,10){$-$}
\put(-82,102){$+$}
\put(-59,20){$+$}
\put(-59,92){$-$}
\put(-111,69){$-$}
\put(-111,45){$-$}
\put(-31,69){$+$}
\put(-31,45){$-$}
\put(-75,86){$+$}
\put(-75,27){$+$}
\caption{Internal helicities can be arranged in two valid configurations on 
hepta-cut solution $\mc S_2$ in the $-+-+$ amplitude.}
\label{-+-+S2}
\ec
\end{figure}

We next turn to the $-+-+$ helicity amplitude and work through the contribution
to the master integral coefficients due to hepta-cut solution $\mc S_2$. There
are two possible assignments $A$ and $B$ of helicities on internal lines, shown 
in fig.~\ref{-+-+S2}. 

Again, the product of tree amplitudes is very simple when evaluated in the
maximally supersymmetric theory,
\begin{align}
\sum_{\substack{\mc N = 4\\ \text{multiplet}}}
\prod_{j=1}^6 A_{(j)}^\tree(z)\big|_{\mc S_i} = -s_{12}^2s_{14}A^\tree_{-+-+}\;,
\end{align}
whence we need to determine the ratio for $\mc N = 2,1,0$ supersymmetries. In 
the case at hand, single $SU(4)$ factors read
\begin{align}
A = {} &
\avg{1p_1}\Avg{p_1p_6}\avg{p_6p_7}\Avg{p_7p_3}\avg{p_33}\Avg{p_54}\;, \\[1mm]
B = {} & 
-\avg{1p_2}\Avg{p_2p_7}\avg{p_7p_6}\Avg{p_6p_5}\avg{3p_4}\Avg{p_44}\;.
\end{align}
The string of spinor products in the denominator is of course still given by 
\eqref{S2CUTDEN} as in the previous example. By multiple applications of 
momentum conservation and insertion of the explicit hepta-cut solutions,
\begin{align}
p_3^\mu = \ell_2^\mu-k_4^\mu = 
\frac{s_{12}z}{2\avg{41}\Avg{13}}\expval{4^-}{\gamma^\mu}{3^-}+k_3^\mu\;, \\
p_5^\mu = \ell_2^\mu+k_3^\mu = 
\frac{s_{12}z}{2\avg{41}\Avg{13}}\expval{4^-}{\gamma^\mu}{3^-}-k_4^\mu\;,
\end{align}
it is not hard to realize that
\begin{align}
\frac{A}{A+B} = \frac{\aAvg{1}{p_5}{4}}{\aAvg{1}{2}{4}} = \frac{z}{1+\chi}\;.
\end{align}
The generically supersymmetric state sum becomes
\begin{align}
\sum_{\substack{\mc N\leq 4\\ \text{multiplet}}}
\prod_{j=1}^6 A_{(j)}^\tree(z)\big|_{\mc S_2} = 
-s_{12}^2s_{14}A^\tree_{-+-+}\left\{1+(4-\mc N)
\frac{z(z-\chi-1)}{(1+\chi)^2}\right\}\;.
\end{align}
Then we finally plug the supersymmetric sum together with contributions from 
the other on-shell branches, which we do not include explicitly, into the master 
integral formulae and derive the alternating helicity amplitude
\begin{align}
A^\text{xbox}_{-+-+} =  -s_{12}^2s_{14}A^\tree_{-+-+}\bigg\{
\left(1+(4-\mc N)\frac{s_{14}}{4s_{13}}\right)\INP[1] \hspace*{4.5cm} \nn \\
+(4-\mc N)\frac{s_{13}+3s_{14}}{2s_{13}^2}\INP[(\ell_1\cdot k_3)]\bigg\}\;,
\end{align}
where the coefficients are valid to $\mc O(\epsilon^0)$.

\section{Integrand-Level Reduction Methods}
Recently other promising methods for two-loop amplitudes such as integrand basis 
determination by multivariate polynomial division algorithms using Gr\"{o}bner 
bases and classification of on-shell solutions by primary decomposition based on 
computational algebraic geometry have been reported \cite{Mastrolia:2011pr,
Badger:2012dv,Zhang:2012ce,Feng:2012bm,Mastrolia:2012an,Mastrolia:2012wf,
Mastrolia:2012du,Kleiss:2012yv,Huang:2013kh}. 

In particular, using hepta-cuts, Gram matrix relations and polynomial fitting 
techniques Badger, Frellesvig and Zhang \cite{Badger:2012dp} were able to obtain 
master integral coefficients in any renormalizable four-dimensional gauge theory 
for the planar double box, two-loop crossed box and pentabox-triangle primitive 
amplitudes, although it turns out that the latter is reducible to simpler 
topologies that contribute to hexacuts for example. In this section we provide 
a very brief review of their method and a comparison to that of Kosower and 
Larsen used here.

\subsection{Irreducible Integrand Bases}
Let $\ell_1$ and $\ell_2$ be the loop momenta and that suppose 
$\{e_1,e_2,e_3,e_4\}$ spans the space of 4-dimensional momenta, say three 
external momenta $\{k_1,k_2,k_4\}$ supplemented with a spurious vector that is
orthogonal to those directions and satisfies $\omega^2 > 0$. Eliminating 
contractions that are trivially reducible using proper combinations of inverse 
propagators and constant terms such as for instance
\begin{align}
2(\ell_1\cdot k_1) = {} & (\ell_1-k_1)^2-\ell_1^2-k_1^2\;, \nn \\
2(\ell_2\cdot k_3) = {} & (\ell_2-k_3)^2-(\ell_2-k_3-k_4)^2+2k_3\cdot k_4+k_4^2\;,
\end{align}
a completely general integrand can be parametrized with four irreducible scalar 
products
\begin{align}
\{\ell_1\cdot k_4,\,\ell_2\cdot
k_1,\,\ell_1\cdot\omega,\,\ell_2\cdot\omega\}\;.
\end{align}

Relations from Gram matrix determinants impose further nontrivial constraints on 
the general form of the integrand. To motivate this we first define for $2n$ 
vectors $\{l_1,\dots,l_n\}$ and $\{v_1,\dots,v_n\}$ the $n\times n$ Gram matrix
by
\begin{align}
G = G\bigg(
\begin{array}{c}
l_1,\dots,l_n \\
v_1,\dots,v_n
\end{array}
\bigg)\;, \quad
G_{ij} = l_i\cdot v_j\;.
\end{align}
We will frequently encounter Gram matrices where the two sets are identical.
Important properties of the Gram determinant $\det G$ are linearity and 
antisymmetry in the vectors in each row. However the real use owes to the fact 
that $\det G$ vanishes if and only if the vectors $\{l_1,\dots,l_n\}$ or 
$\{v_1,\dots,v_n\}$ are linearly dependent. Then if $\ell_1$ and $\ell_2$ are 
also four-dimensional, by this property,
\begin{align}
\det G\bigg(
\begin{array}{c}
e_1,e_2,e_3,\ell_1 \\
e_1,e_2,e_3,\ell_1
\end{array}
\bigg) =
\det G\bigg(
\begin{array}{c}
e_1,e_2,e_3,\ell_2 \\
e_1,e_2,e_3,\ell_2
\end{array}
\bigg) =
\det G\bigg(
\begin{array}{c}
e_1,e_2,e_3,\ell_1 \\
e_1,e_2,e_3,\ell_2
\end{array}
\bigg) = 0\;.
\end{align}
These relations imply that $(\ell_1\cdot\omega)^2$, $(\ell_2\cdot\omega)^2$ and
$(\ell_1\cdot\omega)(\ell_2\cdot\omega)$ are reducible. Other Gram matrix
relations may be derived from combinations of the fundamental three to provide 
additional constraints on the integrand reducing the number of irreducible scalar 
products monomials to 32 and 38 for the planar and nonplanar double box
respectively. We leave the precise summation ranges implicit and write
\begin{align}
\mc N^{\text{P}}(\ell_1,\ell_2) = {} & 
\sum_{m,n,\alpha,\beta}
c_{mn(\alpha+2\beta)}(\ell_1\cdot k_4)^m(\ell_2\cdot k_1)^n
(\ell_1\cdot\omega)^\alpha(\ell_2\cdot\omega)^\beta\;.
\end{align}
The crossed box is similar,
\begin{align}
\mc N^{\text{NP}}(\ell_1,\ell_2) = {} & 
\sum_{m,n,\alpha,\beta}
c_{mn(\alpha+2\beta)}(\ell_1\cdot k_3)^m(\ell_2\cdot k_2)^n
(\ell_1\cdot\omega)^\alpha(\ell_2\cdot\omega)^\beta\;.
\end{align}

\subsection{Master Integral Coefficients}
Badger, Frellesvig and Zhang use the well-known parametrization \eqref{LOOPPARAM} 
to solve the equations for the hepta-cut, but without normalizations formed by 
spinor products and momentum invariants in the cross terms. Moreover, the choice of 
momentum flow and the free parameter differ slightly from ours. Using their 
parametrization of the two-loop momenta a general form of the integrand at the 
hepta-cut may be inferred. The cut crossed box has a very simple polynomial form 
for all on-shell solutions,
\begin{align}
\sum_{\substack{\text{helicities}\\\text{particles}}}
\prod_{j=1}^6 A^\tree_{(j)}(\tau) = 
\begin{cases}
\,\sum_{n=0}^6 d_{s,n}\tau^x & s = 1,2,5,6\,, \\[1mm]
\,\sum_{n=0}^4 d_{s,n}\tau^x & s = 3,4,7,8\,,
\end{cases}
\end{align}
while the cut planar double box due to poles in tensor integrals also contains
terms with inverse powers of the free parameter,
\begin{align}
\sum_{\substack{\text{helicities}\\\text{particles}}}
\prod_{j=1}^6 A^\tree_{(j)}(\tau) = 
\begin{cases}
\,\sum_{n=0}^4 d_{s,n} \tau^x & s = 1,2,3,4\,, \\[1mm]
\,\sum_{n=-4}^4 d_{s,n} \tau^x & s = 5,6\,.
\end{cases}
\end{align}

Schematically it is now possible to construct a $48\times 38$ matrix $M$ for the
nonplanar double box relating the coefficients in the integrand to those in the 
product of tree amplitudes such that
\begin{align}
\mathbf{d} = M\mathbf{c} \;\LLra\; \mathbf{c} = (M^TM)^{-1}M^T\mathbf{d}\;,
\end{align}
whereas the matrix in the case of the planar double box has dimensions
$38\times 32$. The matrix $M$ has full rank and analytical inversion of the 
hepta-cut matrix equations and subsequent reduction onto master integrals using the 
integration-by-parts identities produce the following coefficients for the 
nonplanar crossed box, 
\begin{align}
c_1 = {} &
c_{000}+\frac{1}{16}s_{14}s_{13}(c_{200}-c_{110}+2c_{020})          
\nn \\[1mm] {} &
+\frac{1}{32}s_{14}s_{13}(s_{14}-s_{13})
(c_{300}-c_{210}+c_{120}-2c_{030})
\nn \\[1mm] {} &
+\frac{1}{16^2}(3(s_{14}-s_{13})^2+s_{12}^2)s_{14}s_{13}(
c_{400}-c_{310}+c_{220}+2c_{040}) \nn \\[1mm] {} &
+\frac{1}{16^2}((s_{14}-s_{13})^2+s_{12}^2)s_{14}s_{13}(s_{14}-s_{13})(
c_{320}-c_{410}-2c_{050}) \nn \\[1mm] {} &
+\frac{1}{16^3}(5 (s_{14}-s_{13})^4+10s_{12}^2(s_{14}-s_{13})^2+
s_{12}^4)s_{14}s_{13}(c_{420}+2c_{060})\;, \\[3mm]
c_2 = {} & c_{100}-2 c_{010}+\frac{3}{8}(s_{14}-s_{13})(
c_{200}-c_{110}+2c_{020}) \nn \\[1mm] {} &
+\frac{1}{16}(2 (s_{14}-s_{13})^2+s_{12}^2)(
c_{300}-c_{210}+c_{120}-2c_{030}) \nn \\[1mm] {} &
+\frac{2}{16^2}(5(s_{14}-s_{13})^2+7s_{12}^2)(s_{14}-s_{13})(
c_{400}-c_{310}+c_{220}+2c_{040}) \nn \\[1mm] {} &
+\frac{1}{16^2}(3 (s_{14}-s_{13})^4+8s_{12}^2(s_{14}-s_{13})^2+
s_{12}^4)(c_{320}-c_{410}-2c_{050}) \nn \\[1mm] {} &
+\frac{2}{16^3}\left(7(s_{14}-s_{13})^4+30
s_{12}^2(s_{14}-s_{13})^2+11s_{12}^4\right)
(s_{14}-s_{13})(c_{420}+2c_{060})\;,
\end{align}
and for the planar double box,
\begin{align}
c_1 = {} &
c_{000}+\frac{s_{12}s_{14}}{8}c_{110}
-\frac{s_{12}^2s_{14}}{16}\left(c_{120}+c_{210}\right)
+\frac{s_{12}^3s_{14}}{32}\left(c_{130}+c_{310}\right) \nn \\[1mm] {} &
-\frac{s_{12}^4s_{14}}{64}\left(c_{140}+c_{410}\right)\;, \\[3mm]
c_2 = {} & c_{100}+c_{010}-\frac{3s_{12}}{4}c_{110}
+\frac{s_{14}}{2}\left(c_{020}+c_{200}\right)
+\frac{3s_{12}^2}{8}\left(c_{120}+c_{210}\right) \nn \\[1mm] {} &
+\frac{s_{14}^2}{4}\left(c_{030}+c_{300}\right)
-\frac{3s_{12}^3}{16}\left(c_{130}+c_{310}\right)
+\frac{s_{14}^3}{8}\left(c_{040}+c_{400}\right) \nn \\[1mm] {} &
+\frac{3s_{12}^4}{32}\left(c_{140}+c_{410}\right)\;.
\end{align} 
Several additional null-space conditions are generated in the process. The structure 
of these is equivalent to the redundancy of global residues identified previously.

In order to unify the two approaches we have to synchronize the
parametrizations of the loop momenta on the hepta-cut. Refer to \cite{Badger:2012dp} 
for the explicit parameters and conventions for the free parameter $\tau$. It can 
be shown that this is achieved for the crossed box if
\begin{align}
\mc S_i\,:\;\; \tau(z) = -s_{12}z\;,
\end{align}
which in particular means that the Jacobians remain simple and uniform across all 
on-shell solutions. As a consequence of the the additional poles in tensor 
integrals the situation is a bit more complicated for the planar double box. We 
find that the two parametrizations agree everywhere on the hepta-cut if
\begin{align}
\mc S_1,\dots,\mc S_4:\; \tau(z) = 1+\frac{z}{\chi}\;, \;\;\quad
\mc S_5,\mc S_6:\; \tau(z) = -\frac{1}{z+\chi+1}\;.
\end{align}
with the interchange $\mc S_2 \longleftrightarrow S_3$ due to the fact that
solutions are labeled differently. We can now apply the displayed 
transformations to the master formulae, carefully keeping track of extra Jacobian 
factors and how the global poles are mapped. For instance we see that poles in 
tensor integrands at $z = -\chi-1$ are shifted to $\tau = \infty$. The contour 
weights are of course not affected. After all the master integral coefficients for 
the planar double box can be written
\begin{align}
c_1 = {} & +\frac{1}{4}\sum_{i=1,3}\Res_{\tau=0}\frac{1}{\tau}
\sum_{\substack{\text{particles}\\ \text{helicities}}}
\prod_{j=1}^6 A_{(j)}^\tree(z)\big|_{\mc S_i} \nn \\ &
+\frac{1}{4}\sum_{i=5,6}\Res_{\tau=-1}\frac{1}{1+\tau}
\sum_{\substack{\text{particles}\\ \text{helicities}}}
\prod_{j=1}^6 A_{(j)}^\tree(\tau)\big|_{\mc S_i} \nn \\ &
-\frac{\chi}{4}\sum_{i=5,6}\Res_{\tau=\infty}
\frac{1}{(1+\tau)(1+(1+\chi)\tau)}
\sum_{\substack{\text{particles}\\ \text{helicities}}}
\prod_{j=1}^6 A_{(j)}^\tree(\tau)\big|_{\mc S_i}\;, \\[2mm]
c_2 = {} & 
-\frac{1}{2s_{12}\chi}\sum_{i=1,3}
\Res_{\tau=0}\frac{1}{\tau}
\sum_{\substack{\text{particles}\\ \text{helicities}}}
\prod_{j=1}^6 A_{(j)}^\tree(z)\big|_{\mc S_i} \nn \\ &
+\frac{1+\chi}{s_{12}\chi}\sum_{i=5,6}
\Res_{\tau=-\frac{1}{1+\chi}}\frac{1}{1+(1+\chi)\tau}
\sum_{\substack{\text{particles}\\ \text{helicities}}}
\prod_{j=1}^6 A_{(j)}^\tree(\tau)\big|_{\mc S_i} \nn \\ &
-\frac{1}{2s_{12}\chi}\sum_{i=5,6}
\Res_{\tau=-1}\frac{1}{1+\tau}
\sum_{\substack{\text{particles}\\ \text{helicities}}}
\prod_{j=1}^6 A_{(j)}^\tree(\tau)\big|_{\mc S_i} \nn \\ &
+\frac{3}{2s_{12}}\sum_{i=5,6}
\Res_{\tau=\infty}
\frac{1}{(1+\tau)(1+(1+\chi)\tau)}
\sum_{\substack{\text{particles}\\ \text{helicities}}}
\prod_{j=1}^6 A_{(j)}^\tree(\tau)\big|_{\mc S_i}\;.
\end{align}
Along these lines it is straightforward to obtain expressions for the master integral 
coefficients formulated in terms of residues that are compatible with any 
parametrization of the loop momenta.

The explicit mapping between the integrand basis coefficients and the tree-level data
is quite complicated. Using Mathematica we are able shuffle around the null-space 
conditions appropriately in order to establish full analytical equivalence between the 
two master integral coefficients for both the planar and nonplanar double box prior to 
any reference to particle content of the gauge theory in consideration.

\section{Conclusion}
The unitarity method has been applied widely with great success to otherwise
unattainable computations of loop corrections to scattering amplitudes. In particular, 
generalized unitarity provides means for determining one-loop amplitudes from an 
integral basis whose elements are known explicitly. By imposing multiple simultaneous 
on-shell conditions on internal propagators, single integrals are projected and their
coefficients are expressed in terms of tree-level data. 

In this paper we have extended four-dimensional maximal unitarity
\cite{Kosower:2011ty} to the nonplanar case. In maximal unitarity computations all 
propagators are cut by placing them on their mass-shell. The massless four-point 
two-loop nonplanar double box admits expansion onto two master integrals that are 
sensitive to hepta-cuts. Maximal cuts are naturally defined by promoting real slice 
Feynman integrals to multidimensional complex contour integrals encircling the global 
poles of the loop integrand while requiring continued validity of all integral 
reduction identities. In order to conform with this principle, each global pole or 
contour must have a weight. We used this approach to derive unique and strikingly 
compact formulae for both master integral coefficients. Moreover, we compared our
results to coefficients recently computed by integrand-level reduction and found 
exact agreement in any renormalizable gauge theory with adjoint matter.

We finally mention several interesting directions for future research. It would of 
course be extremely useful to have formulae for master integral coefficients for 
subleading topologies. However, these integrals are only accessible using cuts with 
fewer propagators and thus more complicated to isolate. It is certainly also
important to consider $D$-dimensional unitarity cuts in order to capture pieces
that are not detectable in four dimensions. Indeed, it is possible to establish
nonzero linear combinations of tensor integrals whose hepta-cuts vanish
identically at $\mc O(\epsilon^0)$ \cite{Kosower:2011ty}. However, the most urgent 
point to address is probably how integration-by-parts identities constrain 
contours. In particular, a deeper understanding of the unexpected simplicity of 
contour weights is desirable. A very natural extension of our work is to study 
nonplanar double boxes with one or more massive external legs or even internal 
masses. Guided by recent results for planar triple boxes obtained by integrand 
reconstruction we also expect that the framework of maximal unitarity can be 
applied beyond two loops. We hope to return to some of these questions soon.

\acknowledgments
It is a pleasure to thank Emil Bjerrum-Bohr, Poul Henrik Damgaard and Yang Zhang 
for many stimulating discussions. The author is grateful to the theoretical 
elementary particle physics group at UCLA and in particular Zvi Bern for 
hospitality during the completion of this work.

\appendix
\clearpage
\section{Kinematical Configurations of the Two-Loop Crossed Box}
We depict here the eight valid kinematical configurations of the maximally cut
twoloop crossed box, labelled according to solutions $\mc S_1,\dots,\mc S_8$. 
Holomorphically-collinear and antiholomorphically-collinear three-vertices are 
represented by $\ominus$ and $\oplus$ respectively.
\begin{figure}[!h]
\bc
\includegraphics[scale=0.6]{holo1} \hspace*{0.5cm}
\put(-17,57){$k_1$}
\put(-165,57){$k_2$}
\put(-220,0){$k_3$}
\put(-220,115){$k_4$}
\put(-140,115){$\mc S_1$}
\includegraphics[scale=0.6]{holo2} 
\put(2,57){$k_1$}
\put(-148,57){$k_2$}
\put(-201,0){$k_3$}
\put(-201,115){$k_4$} 
\put(-121,115){$\mc S_2$} \\
\vspace*{.5cm}
\includegraphics[scale=0.6]{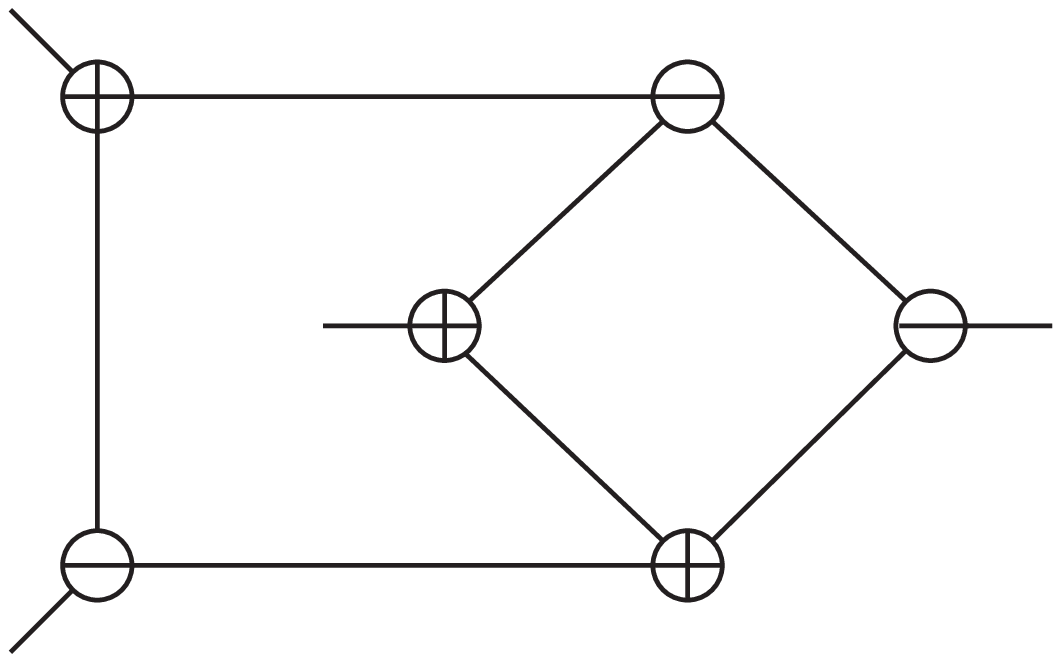} \hspace*{0.5cm}
\put(-17,57){$k_1$}
\put(-165,57){$k_2$}
\put(-220,0){$k_3$}
\put(-220,115){$k_4$}
\put(-140,115){$\mc S_3$}
\includegraphics[scale=0.6]{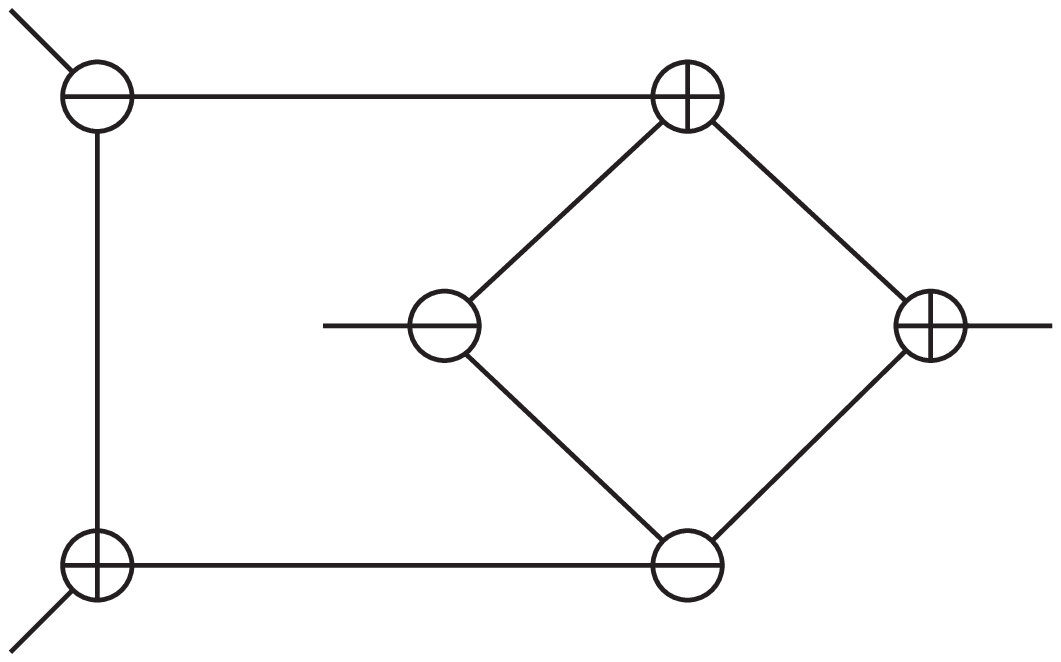}
\put(2,57){$k_1$}
\put(-148,57){$k_2$}
\put(-201,0){$k_3$}
\put(-201,115){$k_4$}
\put(-121,115){$\mc S_4$} \\
\vspace*{.5cm}
\includegraphics[scale=0.6]{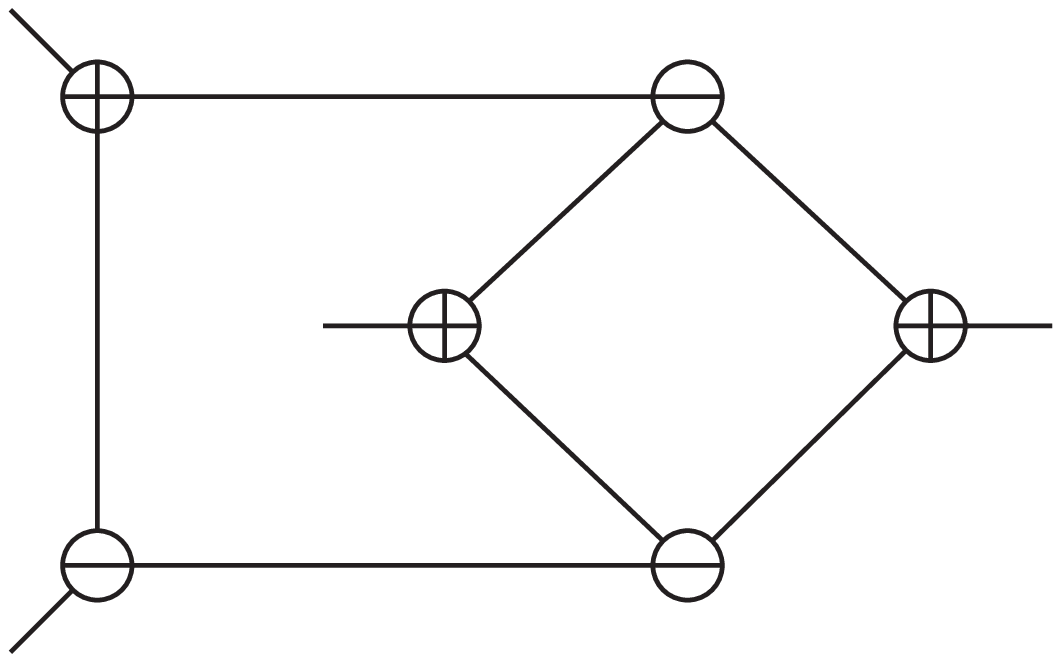} \hspace*{0.5cm}
\put(-17,57){$k_1$}
\put(-165,57){$k_2$}
\put(-220,0){$k_3$}
\put(-220,115){$k_4$}
\put(-140,115){$\mc S_5$}
\includegraphics[scale=0.6]{holo6}
\put(2,57){$k_1$}
\put(-148,57){$k_2$}
\put(-201,0){$k_3$}
\put(-201,115){$k_4$}
\put(-121,115){$\mc S_6$} \\
\vspace*{.5cm}
\includegraphics[scale=0.6]{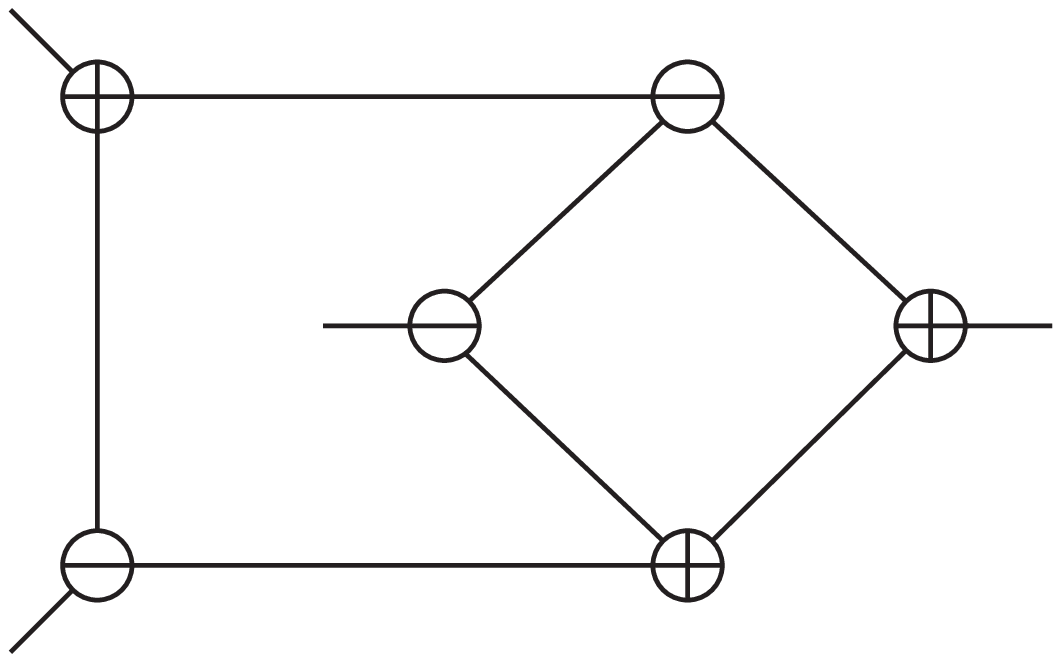} \hspace*{0.5cm}
\put(-17,57){$k_1$}
\put(-165,57){$k_2$}
\put(-220,0){$k_3$}
\put(-220,115){$k_4$}
\put(-140,115){$\mc S_7$}
\includegraphics[scale=0.6]{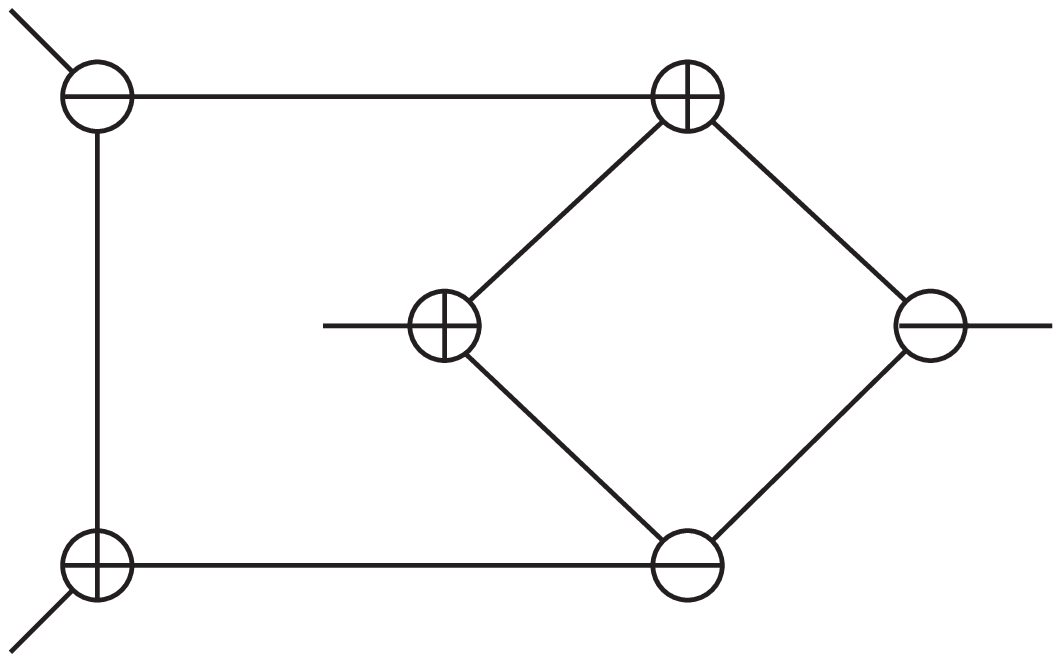}
\put(2,57){$k_1$}
\put(-148,57){$k_2$}
\put(-201,0){$k_3$}
\put(-201,115){$k_4$}
\put(-121,115){$\mc S_8$}
\ec
\label{CROSSEDBOXTREES}
\end{figure}

\clearpage
\section{Two-Loop Crossed Box Integration-By-Parts Identities}
\label{IBPXBOX}
We provide below all four-dimensional integration-by-parts identities used for 
the reduction onto master integrals of all renormalizable four-point tensor 
integrals with two-loop crossed box topology. Ellipses denote truncation at the 
maximal number of propagators.
\begin{align*}
\INP[(\ell_1\cdot k_3)^2] = {} & 
-\frac{1}{16}(1+\chi)\chi s_{12}^2 \INP[1]
+\frac{3}{8}(1+2\chi)s_{12}
\INP[(\ell_1\cdot k_3)]+\cdots \\[2mm]
\INP[(\ell_1\cdot k_3)^3] = {} & 
-\frac{1}{32}\chi(1+\chi)(1+2\chi)s_{12}^3 \INP[1] \\[2mm] &
+\frac{1}{16}(3+8\chi(1+\chi))s_{12}^2
\INP[(\ell_1\cdot k_3)]+\cdots \\[2mm]
\INP[(\ell_1\cdot k_3)^4] = {} &
-\frac{1}{64}\chi(1+\chi)(1+3\chi(1+\chi))s_{12}^4 \INP[1]
\\[2mm] & 
+\frac{1}{32}(1+2\chi)(3+5\chi(1+\chi))s_{12}^3
\INP[(\ell_1\cdot k_3)]+\cdots \\[4mm]
\INP[(\ell_2\cdot k_2)] = {} & 
-2\INP[(\ell_1\cdot k_3)]+\cdots \\[2mm]
\INP[(\ell_2\cdot k_2)^2] = {} & 
-\frac{1}{8}\chi s_{12}^2(1+\chi) \INP[1]
+\frac{3}{4}(1+2\chi)s_{12}
\INP[(\ell_1\cdot k_3)]+\cdots \\[2mm]
\INP[(\ell_2\cdot k_2)^3] = {} & 
+\frac{1}{16}\chi s_{12}^3(1+\chi)(1+2\chi) \INP[1] \\[2mm] &
-\frac{1}{8}s_{12}^2(1+2(1+2\chi)^2)
\INP[(\ell_1\cdot k_3)]+\cdots \\[2mm]
\INP[(\ell_2\cdot k_2)^4] = {} & 
-\frac{1}{128}\chi s_{12}^4(1+\chi)(1+3(1+2\chi)^2) \INP[1]
\\[2mm] & 
+\frac{1}{64}s_{12}^3(1+2\chi)(7s_{12}^2+5(1+2\chi)^2)
\INP[(\ell_1\cdot k_3)]+\cdots \\[2mm]
\INP[(\ell_2\cdot k_2)^5] = {} & 
\frac{1}{128}\chi s_{12}^5(1+\chi)(1+2\chi)(1+(1+2\chi)^2) \INP[1]
\\[2mm] & 
-\frac{1}{128}s_{12}^4(1+(1+2\chi)^2(8+3(1+2\chi)^2))
\INP[(\ell_1\cdot k_3)]+\cdots \\[2mm]
\INP[(\ell_2\cdot k_2)^6] = {} & 
-\frac{1}{2048}\chi s_{12}^6(1+\chi)(1+(1+2\chi)^2(10+5(1+2\chi)^2)) \mc
\INP[1]
\\[2mm] & 
+\frac{1}{1024}s_{12}^5(1+2\chi)(11+(1+2\chi)^2(30+7(1+2\chi)^2))
\INP[(\ell_1\cdot k_3)]+\cdots \\[4mm]
\INP[(\ell_1\cdot k_3)(\ell_2\cdot k_2)] = {} &
+\frac{1}{16}\chi(1+\chi)s_{12}^2 \INP[1]
-\frac{3}{8}(1+2\chi)s_{12}
\INP[(\ell_1\cdot k_3)]+\cdots \\[2mm]
\INP[(\ell_1\cdot k_3)^2(\ell_2\cdot k_2)] = {} &
+\frac{1}{32}\chi(1+\chi)(1+2\chi)s_{12}^3 \INP[1] \\[2mm] &
-\frac{1}{16}(3+8\chi(1+\chi))s_{12}^2
\INP[(\ell_1\cdot k_3)]+\cdots
\end{align*}
\begin{align*}
\INP[(\ell_1\cdot k_3)^3(\ell_2\cdot k_2)] = {} &
+\frac{1}{64}\chi(1+\chi)(1+3\chi(1+\chi))s_{12}^4 \INP[1]
\\[2mm] & 
-\frac{1}{32}(1+2\chi)(3+5\chi(1+\chi))s_{12}^3
\INP[(\ell_1\cdot k_3)]+\cdots \\[2mm]
\INP[(\ell_1\cdot k_3)^4(\ell_2\cdot k_2)] = {} &
+\frac{1}{128}\chi(1+\chi)(1+2\chi)(1+2\chi(1+\chi))s_{12}^5 \INP[1]
\\[2mm] & 
-\frac{1}{256}s_{12}^4(1+(1+2\chi)^2(8+3(1+2\chi)^2))
\INP[(\ell_1\cdot k_3)]+\cdots \\[4mm]
\INP[(\ell_1\cdot k_3)(\ell_2\cdot k_2)^2] = {} &
-\frac{1}{32}\chi(1+\chi)(1+2\chi)s_{12}^3 \INP[1]
\\[2mm] &
+\frac{1}{16}s_{12}^2(3+8\chi(1+\chi)) 
\INP[(\ell_1\cdot k_3)]+\cdots \\[2mm]
\INP[(\ell_1\cdot k_3)^2(\ell_2\cdot k_2)^2] = {} &
-\frac{1}{64}\chi(1+\chi)(1+3\chi(1+\chi))s_{12}^4 \INP[1]
\\[2mm] &
-\frac{1}{64}s_{12}^4\chi(1+\chi)(1+3\chi(1+\chi)) 
\INP[(\ell_1\cdot k_3)]+\cdots \\[2mm]
\INP[(\ell_1\cdot k_3)^3(\ell_2\cdot k_2)^2] = {} &
-\frac{1}{128}\chi(1+\chi)(1+2\chi)(1+2\chi(1+\chi))s_{12}^5 \INP[1]
\\[2mm] & 
+\frac{1}{256}s_{12}^4(1+(1+2\chi)^2(8+3(1+2\chi)^2))
\INP[(\ell_1\cdot k_3)]+\cdots \\[2mm]
\INP[(\ell_1\cdot k_3)^4(\ell_2\cdot k_2)^2] = {} &
-\frac{1}{4096}s_{12}^6\chi(1+\chi)(1+(1+2\chi)^2(10+5(1+2\chi)^2)) \mc
\INP[1]
\\[2mm] & 
+\frac{1}{2048}s_{12}^5(1+2\chi)(11+(1+2\chi)^2(30+7(1+2\chi)^2))
\INP[(\ell_1\cdot k_3)]+\cdots 
\end{align*}

\clearpage
\section{Planar Double Box Integration-By-Parts Identities}
\label{IBPDBOX}
For completeness we also include all truncated integration-by-parts identities
relevant for the planar double box with four massless external lines.
\begin{align*}
\IP[(\ell_1\cdot k_4)^2] = {} & 
\frac{1}{2}\chi s_{12}
\IP[(\ell_1\cdot k_4)]+\cdots \\[2mm]
\IP[(\ell_1\cdot k_4)^3] = {} & 
\frac{1}{4}\chi^2 s_{12}^2
\IP[(\ell_1\cdot k_4)]+\cdots \\[2mm]
\IP[(\ell_1\cdot k_4)^4] = {} &
\frac{1}{8}\chi^3s_{12}^3
\IP[(\ell_1\cdot k_4)]+\cdots \\[4mm]
\IP[(\ell_2\cdot k_1)] = {} & 
\IP[(\ell_1\cdot k_4)]+\cdots \\[2mm]
\IP[(\ell_2\cdot k_1)^2] = {} & 
\frac{1}{2}\chi s_{12}
\IP[(\ell_1\cdot k_4)]+\cdots \\[2mm]
\IP[(\ell_2\cdot k_1)^3] = {} & 
\frac{1}{4}\chi^2 s_{12}^2
\IP[(\ell_1\cdot k_4)]+\cdots \\[2mm]
\IP[(\ell_2\cdot k_1)^4] = {} & 
\frac{1}{8}\chi^3 s_{12}^3
\IP[(\ell_1\cdot k_4)]+\cdots \\[2mm]
\IP[(\ell_1\cdot k_4)(\ell_2\cdot k_1)] = {} &
\frac{1}{8}\chi s_{12}^2 \IP[1]
-\frac{3}{4}s_{12}\IP[(\ell_1\cdot k_4)]+\cdots \\[2mm]
\IP[(\ell_1\cdot k_4)^2(\ell_2\cdot k_1)] = {} &
-\frac{1}{16}\chi s_{12}^3\IP[1]
+\frac{3}{8}s_{12}^2\IP[(\ell_1\cdot k_4)]+\cdots \\[2mm]
\IP[(\ell_1\cdot k_4)^3(\ell_2\cdot k_1)] = {} &
\frac{1}{32}\chi s_{12}^4\IP[1] 
-\frac{3}{16}s_{12}^3\IP[(\ell_1\cdot k_4)]+\cdots \\[2mm]
\IP[(\ell_1\cdot k_4)^4(\ell_2\cdot k_1)] = {} &
-\frac{1}{64}\chi s_{12}^5 \IP[1] 
+\frac{3}{32}s_{12}^4
\IP[(\ell_1\cdot k_4)]+\cdots \\[4mm]
\IP[(\ell_1\cdot k_4)(\ell_2\cdot k_1)^2] = {} &
-\frac{1}{16}\chi s_{12}^3 \IP[1]
+\frac{3}{8}s_{12}^2\IP[(\ell_1\cdot k_4)]+\cdots \\[2mm]
\IP[(\ell_1\cdot k_4)(\ell_2\cdot k_1)^3] = {} &
\frac{1}{32}\chi s_{12}^4\IP[1]
-\frac{3}{16}s_{12}^3\IP[(\ell_1\cdot k_4)]+\cdots \\[2mm]
\IP[(\ell_1\cdot k_4)(\ell_2\cdot k_1)^4] = {} &
-\frac{1}{64}\chi s_{12}^5\IP[1]
+\frac{3}{32}s_{12}^4\IP[(\ell_1\cdot k_4)]+\cdots \\[2mm]
\end{align*}

\end{document}